\documentclass[journal]{IEEEtran}
\usepackage{color}
\usepackage[pdftex]{graphicx}
\DeclareGraphicsExtensions{.pdf,.jpeg,.png}
\graphicspath{{./Fig/}}

\usepackage[cmex10]{amsmath}
\usepackage{amssymb, latexsym}
\usepackage{amsfonts}
\usepackage{amsthm}
\newcommand{\be}{\begin{equation}}
\newcommand{\ee}{\end{equation}}
\def\bal#1\eal{\begin{align}#1\end{align}}
\renewcommand{\Vec}[1]{\mathbf{#1}}
\newcommand{\Mat}[1]{\mathbf{#1}}
\newcommand{\Exp}[1]{\mathrm{e}^{#1}}
\newcommand{\trace}[1]{\mathop{\mathrm{tr}}\left({#1}\right)}

\newcommand{\A}{\Mat{A}_{\mathcal{M}}}
\newcommand{\G}{\Mat{\Gamma}_{\mathcal{M}}}
\renewcommand{\P}{\Mat{P}}

\newcounter{saveeqn}
\newcommand{\alpheqn}{\stepcounter{equation}%
\setcounter{saveeqn}{\value{equation}}\setcounter{equation}{0}%
\renewcommand{\theequation}
            {\arabic{saveeqn}\alph{equation}}}
\newcommand{\reseteqn}{\setcounter{equation}{\value{saveeqn}}%
\renewcommand{\theequation}{\arabic{equation}}}

\newcommand{\Vecgamma}{\Vec{\gamma}\hspace*{-1.3ex}\Vec{\gamma}\hspace*{-1.3ex}\Vec{\gamma}\hspace*{-1.3ex}\Vec{\gamma}\hspace*{-1.3ex}\Vec{\gamma}}
\newcommand{\mtight}{\setlength{\thickmuskip}{0mu} \setlength{\medmuskip}{0mu} \setlength{\thinmuskip}{0mu}}

\begin{document}

\title{  Sparse Bayesian Learning for DOA Estimation in Heteroscedastic Noise}
%There are wavelengths people cannot see,  sounds people cannot hear, and maybe computers have thoughts that people cannot think.

\author{
Peter Gerstoft \IEEEmembership{Member, IEEE,}
Santosh Nannuru \IEEEmembership{Member, IEEE,}
Christoph F. Mecklenbr\"auker \IEEEmembership{Senior Member,~IEEE,}
and
Geert Leus \IEEEmembership{Fellow, IEEE}
\thanks{P.\ Gerstoft and S.\ Nannuru are with University of California San Diego, La Jolla, CA 92093-0238,
USA, http://noiselab.ucsd.edu
}
\thanks{C.F.\ Mecklenbr\"auker is with Institute of Telecommunications,
Vienna University of Technology,
1040 Vienna, Austria, {\tt cfm@ieee.org}}
%\\*[1ex] Draft manuscript for ITA 2016, version
%\thanks{Copyright (c) 2015 IEEE. Personal use of this material is permitted. However, permission to use this material for any other purposes must be obtained from the IEEE by sending a request to pubs-permissions@ieee
\thanks{
G.\ Leus is with the Dept. of Electrical Eng., Math. and Comp. Science, Delft Univ. of Technology, Delft, The Netherlands, g.j.t.leus@tudelft.nl}
\thanks{Supported by
the Office of Naval Research, Grant Nos. N00014-1110439
and FTW Austria's ``Compressed channel state information feedback for time-variant MIMO channels''.}
}
\maketitle

\begin{abstract}
\today\\
The paper considers direction of arrival (DOA) estimation from long-term observations in a noisy environment.
In such an environment the noise source might evolve, causing the stationary models to fail.
Therefore a heteroscedastic Gaussian noise model is introduced where the variance can vary across observations and sensors.
The  source amplitudes  are  assumed independent zero-mean complex Gaussian distributed with  unknown variances (i.e. the source powers), inspiring  stochastic maximum likelihood DOA estimation.
The DOAs of plane waves are estimated from multi-snapshot sensor array data using sparse Bayesian learning (SBL) where the noise is estimated across both sensors and snapshots.
This SBL approach is more flexible and performs better than high-resolution methods since they cannot estimate the heteroscedastic noise process.
An alternative to SBL is simple data normalization, whereby only the phase across the array is utilized.
Simulations demonstrate that taking the heteroscedastic noise into account  improves DOA estimation.

%We demonstrate this model using array processing to find direction of arrivals
%The plan is first doing the heteroscedastic noise correctly using SBL.\\
%Hopefully we will find room for the phase-only processing as well. That is important for applications.\\
 \end{abstract}
%{\keywords Bayesian estimation, sparsity, relevance vector machine}
% EDICS: SSP-TRAC, SAM-DOAE.
%
% For peer review papers, this IEEEtran command inserts a page break and
\begin{IEEEkeywords}
 Heteroscedastic noise, sparse reconstruction, array processing, DOA estimation, compressive beamforming, phase-only processing
\end{IEEEkeywords}
\IEEEpeerreviewmaketitle

\section{Introduction}
With long observation times  weak signals  can be extracted in a noisy environment.
Most analytic treatments analyze these cases assuming   Gaussian noise with constant variance. For long observation times the noise process though is likely to change with time causing the noise variance to evolve.
This is called a heteroscedastic  Gaussian process, meaning that the noise variance is evolving.
While the noise variance is a nuisance parameter that we are not interested in, it still needs to be estimated or included in the processing in order to obtain an accurate estimate of the weaker signals.

Accounting for the noise variation is certainly important for machine learning~\cite{Murphy2012,Bishop2006} and related to robust statistics~\cite{huber2011,maronna2006}.
Heteroscedastic noise models have been used in e.g. finance~\cite{engle1982} and image processing~\cite{thai2014}.
In statistical signal processing, the noise has been assumed to vary spatially~\cite{viberg1997,chen2008,li2011}, but spatiotemporally varying noise as considered here has not been studied.
The proposed processing could be applied to spatial coherence loss \cite{cox1973line,paulraj1988,lefort2017} or to wavefront decorrelation, where turbulence causes the wave front to be incoherent for certain observations (thus more noisy).
 This has lead to so-called lucky imaging in astronomy~\cite{law2006} or lucky ranging in ocean acoustics~\cite{ge2016}, where only the measurements giving good results are used. 
 As a result an involved hypothesis testing is needed to determine the measurements to be used. In contrast, we propose to use all measurements.

In applications, a simple way to  account for noise power variations is to normalize the data magnitude to only contain phase information as demonstrated for beamforming in seismology~\cite{gerstoft2006,gerstoft2016}, noise cross correlation in seismology ~\cite{roux2005,harmon2008,landes2010,zhan2010,weemstra2014} and acoustics~\cite{gerstoft2008},
source deconvolution in ocean acoustics~\cite{sabra2004,sabra2010} and speaker localization~\cite{schwartz2014,dorfan2015}.  High-resolution beamformers such as MUSIC~\cite{VanTreesBook} relying on a sample covariance matrix are not likely to perform well for heteroscedastic noise as the loud noise might dominate the sample covariance matrix. More advanced methods than an eigenvalue decomposition are needed to separate the signal and noise subspaces.
We demonstrate that for well-separated sources normalizing the data to only contain phase information works well.

When the sources are closely spaced, more advanced parametric methods are needed for DOA estimation when the noise power is varying in space and time and the sources are weak. We derive and demonstrate this for the application of multiple measurement vector (MMV, or multi-snapshot) compressive beamforming~\cite{MalioutovDOA:2005,XenakiCS:2014,Wipf2007, Gerstoft2015}. We solve the MMV problem using the sparse Bayesian learning (SBL) framework~\cite{Wipf2007,gerstoft2016mmv,nannuru2017} and use the maximum-a-posteriori (MAP) estimate for DOA reconstruction. We assume the source signals to jointly follow a zero-mean multivariate complex normal distribution with unknown power levels.  The noise across sensors and snapshots also follows a zero-mean multivariate normal distribution with unknown variances. These assumptions lead to a Gaussian likelihood function.

The corresponding posterior distribution is also Gaussian and already developed SBL approaches solve this well. We base our development on our fast SBL method  \cite{gerstoft2016mmv,nannuru2017} which we augment to estimate noise variances, potentially as many variances as observations.
Standard techniques are based on minimization-majorization~\cite{Stoica2012} and expectation maximization (EM)~\cite{Wipf2007,Wipf2007beam,Wipf2004,Zhang2011,Liu2012, Zhang2016,Giri2016}, though not all estimates work well.
Instead, we estimate the unknown variances using stochastic maximum likelihood~\cite{Boehme1985,Jaffer1988,Stoica-Nehorai1995}, modified to obtain noise estimates even for a single observation.

%Robustness to array imperfections \cite{Weiss2014} and extreme noise distributions \cite{Ollila2015} can be important. We expect that the proposed heteroscedastic models can also reduce sensitivity to either these issues.

%%%%%%%%%%%%%
\subsection{Heteroscedastic noise observation model}
\label{sec:noise_models}
For the $l$th observation snapshot, we assume the linear model
\begin{equation}
\Vec{y}_l=\Mat{A}\Vec{x}_l +\Vec{n}_l,
\label{eq:lin3}
\end{equation}
where the dictionary $\Mat{A}\in\mathbb{C}^{N\times M}$ is constant and known, and the source vector $\Vec{x}_l\in\mathbb{C}^{M}$ contains the physical information of interest.
Further, $\Vec{n}_l\in\mathbb{C}^{N}$ is additive zero-mean circularly symmetric complex Gaussian noise, which is generated from a
\emph{heteroscedastic} Gaussian process
$\Vec{n}_l\sim{\cal CN}(\Vec{n}_l;\Vec{0}, \Mat{\Sigma}_{\Vec{n}_l})$.
Due to the circular symmetry of the noise the phase is uniformly distributed.
%All noise covariance matrices are collected in the 3-way array
%\begin{align}
%{\Mat{\Sigma}}_{\Mat N} = [  \Mat\Sigma_{\Vec{n}_l}| \,l=1,\ldots,L ]
%\in \mathbb{C}^{N\times N\times L}.
%\label{eq:sigma_N}
%\end{align}

We specialize to diagonal covariance matrices, parameterized as
\be
\Mat{\Sigma}_{\Vec{n}_l} = \sum\limits_{n=1}^N \sigma^2_{nl} \Mat{J}_n =  \mathop{\mathrm{diag}}(\sigma^2_{1l},\ldots, \sigma^2_{Nl}),
\label{eq:noise-cov-model-1}
\ee
where $\Mat{J}_n=\mathop{\mathrm{diag}}(\Vec{e}_n)=\Vec{e}_n\Vec{e}_n^T$ with $\Vec{e}_n$ the $n$th standard basis vector.
Note that the covariance matrices $\Mat{\Sigma}_{\Vec{n}_l}$ are varying over the snapshot index $l=1,\ldots, L$ and we introduce the matrix of all noise standard deviations as
\begin{align}
\Mat{V}_{\Mat N} = \left(\begin{array}{ccc}
   \sigma_{11} & \hdots & \sigma_{1L} \\
         \vdots         & \ddots& \vdots \\
   \sigma_{N1} & \hdots & \sigma_{NL}
 \end{array} \right) \in \mathbb{R}_{0+}^{N\times L},
\label{eq:noise-std-params}
\end{align}
where $\mathbb{R}_{0+}$ denotes non-negative real numbers.

We consider three special cases for the a priori knowledge on the noise covariance model (\ref{eq:noise-cov-model-1}) which are expressed as constraints on $\Mat{V}_{\Mat N}$.
\begin{description}
\item[\hspace{-2ex}Case I] We assume wide-sense stationarity of the noise in space and time: $\mtight \sigma_{nl}^2=\sigma^2$. The model is homoscedastic,
\begin{align}
\Mat{V}_{\Mat N} \in \mathcal{V}_{\mathrm{I}} = \{ \Mat{V}\in \mathbb{R}_{0+}^{N\times L} | \Mat{V} = \sigma \Vec{1}_N\Vec{1}_L^T \}.
\end{align}
\item[\hspace{-2ex}Case II] We assume wide-sense stationarity of the noise in space only, i.e., the noise variance for all sensor elements is equal across the array, $\sigma_{nl}^2=\sigma_{0l}^2$ and it varies over snapshots. The noise variance is heteroscedastic in time (across snapshots),
\begin{align}
\!\!\!\!\!\!\Mat{V}_{\Mat N}\! \in\! \mathcal{V}_{\mathrm{II}} = \{ \Mat{V}\!\in \!\mathbb{R}_{0+}^{N\times L} | \Mat{V} = (\sigma_{01} \ldots \sigma_{0L}) \Vec{1}_L^T \!\}.
\end{align}
%\item[IIb] We assume wide-sense stationarity of the noise over snapshots only, i.e., the noise variance for all snapshots is equal, but each sensor element is associated with an individual noise variance, $\sigma_{n,l}^2=\sigma_{n,0}^2$ and it varies over snapshots. The noise variance is heteroscedastic in space (across sensors).
%(We will ignore this case for most part of this paper.)
\item[\hspace{-2ex}Case III] No additional constraints other than  (\ref{eq:noise-std-params}). The noise variance is heteroscedastic across  both time and space (sensors and snapshots.). In this case
$\Mat{V}_{\Mat N}\in\mathcal{V}_{\mathrm{III}}=\mathbb{R}_{0+}^{N\times L}$.
\end{description}
The relation between these noise cases is 
$\mtight \mathcal{V}_{\mathrm{I}}\subset\mathcal{V}_{\mathrm{II}}\subset\mathcal{V}_{\mathrm{III}} =\mathbb{R}_{0+}^{N\times L}$.
From the sets $\mathcal{V}_d$ ($d$ is I, II, or III) both $ \Mat{V}_{\Mat N}$ in \eqref{eq:noise-std-params} and $\Mat{\Sigma}_{\Vec{n}_l}$ in  \eqref{eq:noise-cov-model-1} can be constructed. %Noise free data for case I ($\sigma=0$) and II ($\sigma_l=0$) requires special care f
%The sets $\mathcal{V}_d$ are mapped to corresponding sets  $\mathcal{C}_d$ of 3-way arrays expressing the equivalent constraints on ${\Mat{\Sigma}}_{\Mat N}$.

%To be deleted
%\be
%\Mat{\Sigma}_{\Vec{n}_l} = \mathop{\mathrm{diag}}[(\Mat{V}_{\Vec N} \odot \Mat{V}_{\Vec N}) \Vec{e}_l]
% \ee
%where $\Vec{e}_l$ is a unit vector, with the $l$th element being one.
%%%%%%%%%%%%
\subsection{Array model}

Let $\Mat{X}\!\!=\!\![\Vec{x}_1,\ldots,\Vec{x}_L]\!\!\in\!\!\mathbb{C}^{M\times L}$ be the complex source amplitudes, $x_{ml}=[\Mat{X}]_{m,l}=[\Mat{x}_l]_{m}$ with $m\in\{1,\cdots, M\}$ and $l\in\{1,\cdots, L\}$, at $M$ DOAs (e.g., $\theta_m=-90^{\circ}+\frac{m-1}{M}180^{\circ}$) and $L$ snapshots for a frequency $\omega$. We observe narrowband waves on  $N$ sensors for $L$ snapshots
$\Mat{Y}=[\Vec{y}_1,\ldots,\Vec{y}_L]\in\mathbb{C}^{N\times L}$.
A linear regression model relates the  array data
$ \Mat{Y} $  to the source amplitudes  $\Mat{X} $ as
\be
   \Mat{Y} = \Mat{A}\Mat{X} + \Mat{N}.  \label{eq:linear-model3}
 \ee
The dictionary $ \mtight \Mat{A}=[\Vec{a}_1, \ldots,\Vec{a}_M]\in\mathbb{C}^{N\times M}$ contains the array
steering vectors for all hypothetical DOAs as columns, with the $(n,m)$th element given by $\mathrm{e}^{-\mathrm{j}\frac{\omega d_n}{c}\sin\theta_m}$ ($d_n$ is the distance to the reference element  and $c$ the sound speed).

We assume $\mtight M> N$ and thus (\ref{eq:linear-model3}) is underdetermined. In the presence of only few stationary sources, the source vector $\Vec{x}_l$ is  $K$-sparse with $\mtight K\ll M$.
We define the $l$th active set
\be
     \mathcal{M}_l = \{m\in\mathbb{N}| x_{ml}\ne0\},
\label{eq:active-set-l}
\ee
and assume $ \mtight \mathcal{M}_l =   \mathcal{M} = \{m_1,\ldots,\,m_K \}$ is constant across all snapshots $l$.
Also,  we define $\mtight \Mat{A}_{\mathcal{M}}\in\mathbb{C}^{N\times K}$  which contains only the $K$
``active'' columns of $\Mat{A}$.
In the following, $\|\cdot\|_p$  denotes the vector $p$-norm and $\|\cdot\|_{\mathcal{F}}$ the matrix Frobenius norm.

Similar to other DOA estimators, 
$K$ can be estimated by model order selection criteria or by examining the angular spectrum. % ($\Vecgamma$ spectrum).
The  parameter $K$ is required only for  the noise power %$\sigma^2$
 in the SBL algorithm. An inaccurate estimate influences the algorithm's convergence.
 
  %%%%%%%%%%%%%%%%%%%%%
\subsection{Prior on the sources}

%To impose a prior on $\Mat{X}$,
We assume that the complex source amplitudes $x_{ml}$ are independent both across snapshots and across DOAs and follow a zero-mean circularly symmetric complex Gaussian distribution with DOA-dependent variance $\gamma_m $, $ m=1,\dots, M$, %\in \Vecgamma=[\gamma_1,\ldots,\gamma_M]^T$,
\begin{align}
\label{eq:xprior}
    p(x_{ml}; \gamma_m) &= \left\{ \begin{array}{ll}
        \delta(x_{ml}), & \text{for } \gamma_m = 0 \\
        \frac{1}{\pi\gamma_m} \mathrm{e}^{-|x_{ml}|^2/\gamma_m}, & \text{for } \gamma_m > 0
        \end{array} \right. ,
\\
\label{eq:tipping-like}
p(\Mat{X};\Vecgamma) &=  \prod_{l=1}^L \prod_{m=1}^M p(x_{ml}; \gamma_m) %\nonumber \\
=\prod_{l=1}^L {\cal CN}(\Vec{x}_l;\Vec{0},\Mat{\Gamma}),
\end{align}
i.e., the source vector $\Vec{x}_{l}$ at each snapshot $\mtight l\in\{1,\cdots, L\}$ is multivariate Gaussian  with potentially singular covariance matrix,
\be
\Mat{\Gamma} = \mathop{\mathrm{diag}}(\Vecgamma)= \mathop{\mathsf{E}} [   \Vec{x}_l\Vec{x}_l^H; \Vecgamma ],
\ee
as $\mtight \mathop{\mathrm{rank}}(\Mat{\Gamma}) = \mathop{\mathrm{card}}(\mathcal{M}) = K \le M$  (typically $K\ll M$). Note that the diagonal elements of $\mtight \Mat{\Gamma} $, 
denoted as $\mtight\Vecgamma$,  represent source powers and thus ${\Vecgamma} \geq {\bf 0}$. 
When the variance $\mtight \gamma_{m}=0$, then $\mtight x_{ml}=0$ with probability 1.
The sparsity of the model is thus controlled with the parameter $\Vecgamma$,
and the active set $\mathcal{M}$ is equivalently  %\eqref{eq:active-set-gamma}.
\be
     \mathcal{M} = \{m\in\mathbb{N} | \gamma_m > 0 \} \label{eq:active-set-gamma}~.
\ee
The  SBL algorithm ultimately estimates  $\Vecgamma$ rather than the complex source amplitudes $\Mat{X}$. This amounts to a significant reduction of the degrees of freedom resulting in a low variance of the DOA estimates.

%%%%%%%%%%%%%%%%%%%%%%%%%%%%%
\subsection{Stochastic likelihood}

We here derive the well-known stochastic maximum likelihood function ~\cite{Bohme1986,Krim1996,Stoica1996}.
Given the linear model \eqref{eq:linear-model3} with Gaussian source \eqref{eq:tipping-like} and noise \eqref{eq:noise-cov-model-1}
the array data $\Mat{Y}$ is Gaussian with for each snapshot $l$ the covariance $\Mat\Sigma_{\Vec{y}_l}$ given by%and its inverse
\begin{align}
  \Mat\Sigma_{\Vec{y}_l}&=  \mathop{\mathsf{E}} [ \Vec{y}_l \Vec{y}_l^H ]  = \Mat{\Sigma}_{\Mat{n}_l}+  \Mat{A} \Mat{\Gamma}\Mat{A}^H   \label{eq:array-covariance}
 \end{align}
  The probability density function of ${\bf Y}$ is thus given by
\begin{align}
p(\Mat{Y} ) &  =\prod_{l=1}^L {\cal CN}(\Vec{y}_l;\Vec{0},\Mat\Sigma_{\Vec{y}_l})
 = \prod_{l=1}^L\frac{\Exp{-\Mat{y}_l^H \Mat\Sigma_{\Vec{y}_l}^{-1}\ \Mat{y}_l}}{\pi^{N}\det\Mat\Sigma_{\Vec{y}_l}}~. \label{eq:stochastic-likelihood}
  \end{align}
The $L$-snapshot log-likelihood for estimating ${\Vecgamma}$ and $\Mat{V}_{\Mat N}$ is 
% $\{ \sigma \}$ for Noise Case I,  $\{ \Mat\Sigma_{\Vec{n}_l}| l=1,\ldots,L   \}$ for Noise Case II, and  $\{ \Mat\Sigma_{\Vec{n}_{nl}}| n=1,\ldots,N l=1,\ldots,L   \}$ 
%= $\{ \Mat\Sigma_{\Vec{n}_1}, \dots, \Mat\Sigma_{\Vec{n}_L}   \}$ is
\begin{align}
\log &p(\Mat{Y} ; \Vecgamma, \Mat{V}_{\Mat N})
    \propto -\sum_{l=1}^L\left(\Mat{y}_l^H \Mat\Sigma_{\Vec{y}_l}^{-1} \Mat{y}_l +  \log\det\Mat\Sigma_{\Vec{y}_l} \right). %\nonumber
%    &\propto -\trace{\Mat\Sigma_{\Vec{y}}^{-1} \Mat{S}_{\Vec{y}}} - \log\det\Mat\Sigma_{\Vec{y}},
    \label{eq:evidence-L}
\end{align}
This likelihood function is identical to the Type~II likelihood function (evidence) in standard SBL~\cite{Wipf2004,Wipf2007beam,gerstoft2016mmv} which is obtained by treating $\Vecgamma$ as a hyperparameter. The Type~II likelihood is obtained by integrating the likelihood function over the complex source amplitudes, cf.\ (29) in  \cite{gerstoft2016mmv}. The stochastic maximum likelihood approach is used here as it is more direct.

The  parameter estimates $\hat{\Vecgamma}$ and $\widehat{\Mat{V}}_{\Mat N}$ are obtained by maximizing the likelihood, leading to
\be
\label{eq:TypeIIMaxLikelihood}
   (\hat{\Vecgamma},\widehat{\Mat{V}}_{\Mat N}) = \mathop{\arg\max}_{\Vec{\gamma}\ge0,\;\Mat{V}_{\Mat N}\in\mathcal{V}_d} \log  p(\Mat{Y} ; \Vecgamma, \Mat{V}_{\Mat N}),
\ee
where $\mathcal{V}_d$ is the feasible set of noise variances   $ \Mat{V}_{\Mat N}$ in \eqref{eq:noise-std-params} corresponding to  the  noise cases ($d=$ I, II, or III, see Sec.~\ref{sec:noise_models}).%where the convex cone $\mathcal{C}_d$ defines the feasible set of noise covariance matrices corresponding to the prior knowledge on the noise (three cases, $d=$ I, II, or III, as described in Sec.~\ref{sec:noise_models}).
The likelihood function \eqref{eq:evidence-L} is similar to the ones derived for SBL and LIKES~\cite{Stoica2012}.
If $\Vecgamma$ and $\Mat{\Sigma}_{{\Mat n}_l}$ are known, then the MAP estimate is the posterior mean  $ \hat{\Mat x}_l^{\mathrm{MAP}} $ and covariance $ \Mat{\Sigma}_{\Vec{x}_l} $~\cite{Tipping2001,gerstoft2016mmv},
\begin{align}
\hat{\Mat x}_l^{\mathrm{MAP}}&= %\Vec{\mu}_{{\Mat x}_l}=
\Mat\Gamma \Mat{A}^H\Mat{\Sigma}_{{\Vec y}_l}^{-1}\Mat{y}_l,
\label{eq:x-map-estimate}
\\
   \Mat{\Sigma}_{\Vec{x}_l}     &=
    \left(  \Mat{A}^H \Mat{\Sigma}_{\Vec{n}_l}^{-1}\Mat{A} + \Mat{\Gamma}^{-1} \right)^{-1}~.  %,\label{eq:Tipping(32)} 
  \label{eq:Sigma_x}
\end{align}	
The diagonal elements of $\Mat\Gamma $, i.e., $\Vecgamma$, control the row-sparsity of $\hat{\Mat x}_l^{\mathrm{MAP}}$ as for $\gamma_m=0$
the corresponding $m$th row of $ \hat{\Mat x}_l^{\mathrm{MAP}}$ becomes $\Vec{0}^T$.

%%%%%
\subsection{LASSSO versus SBL}

Both LASSO and SBL use the linear model \eqref{eq:linear-model3} with complex zero-mean Gaussian random noise
but they differ in the modeling of the source matrix $\Mat{X}$.

The LASSO approach assumes a  priori $\Mat{X}$  random  with uniformly i.i.d.~distributed phase and Laplace-like prior amplitudes,
\begin{align}
p(\mathbf{X})&=p(\mathbf{x}^{\ell_{2}})\propto\exp(-\lVert\mathbf{x}^{\ell_{2}}\rVert_{1}/\nu),   \label{eq:laplacelike}
\\
[ {\bf x}^{\ell_2} ]_n &= \left(\sum_{l=1}^L |  [ {\bf x}_{l} ]_n |^2 \right)^{1/2} . \label{eq:rowsum}
\end{align}
Thus only the summed amplitudes \eqref{eq:rowsum} are Laplacian. The elements in $\Mat X$ are unknown and must be estimated.
The LASSO approach uses the conditional likelihood (Type I) for $p(\Mat{Y}|\Mat{X}$) and applies Bayes rule  with the prior $p(\Mat{X})$ giving the MAP estimate
\begin{equation}
\begin{aligned}
\widehat{\mathbf{X}}&={\arg\max}\; p(\mathbf{Y} | \mathbf{X})p(\mathbf{X})\\
&=\underset{\mathbf{X}\in\mathbb{C}^{N\times L}}{\arg\min}
 \; \lVert \mathbf{Y} - \mathbf{A}\mathbf{X} \rVert_{\mathcal{F}}^{2} + \mu \lVert \mathbf{x}^{\ell_{2}} \rVert_{1}.
\end{aligned}
\label{eq:CS_lasso_MultiSnap}
\end{equation}
The LASSO approach \eqref{eq:CS_lasso_MultiSnap} estimates the realization of $\Mat{x}_l$ for each snapshot $l$.  %(we need ||y-Ax|| )
The number of parameters to be estimated for the LASSO approach grows linearly with the number of snapshots.

The SBL approach on the other hand, assumes that every column of $\Mat{X}$ is random with the same complex zero-mean Gaussian a priori distribution ${\cal CN}(\Vec{x}_l;\Vec{0},\Mat{\Gamma})$.
The SBL approach uses the likelihood $p(\Mat{Y}| \Vecgamma, \Mat{\Sigma}_{\Mat N})$ in \eqref{eq:TypeIIMaxLikelihood} %, see Sec.~\ref{sec:evidence},
and estimates $\Vecgamma$ but not the realization of $\Mat{X}$.
Therefore, the number of  parameters to be estimated for the SBL approach is independent of the number of snapshots. Thus the quality of the estimate improves faster than with LASSO.

For the heteroscedastic noise model, a different covariance matrix could be included in the data fit of \eqref{eq:CS_lasso_MultiSnap}. However, since the number of parameters to be estimated grows with the number of snapshots, this seems less attractive than using SBL.

%%%%
\subsection{Pre-whitening}
The purpose of this section is to motivate the empirical evidence  \cite{gerstoft2006}--\cite{dorfan2015} that phase-only processing might provide improved estimates over using also amplitudes. This is most likely to happen when sources are not closely spaced and at low SNR as demonstrated in the examples, Sec.\ \ref{se:examples}. 
%Further, we also use pre-whitening in SBL when the noise is high and varying across both sensors and snapshots, see Sec.~\ref{sec:multi-dictionary}.

Let us introduce the factorization $ \Mat\Sigma_{\Vec{n}_l}^{-1} = \Mat{W}_l^H\Mat{W}_l$, where $\Mat{W}_l$ is a square and non-singular matrix. For our particular setup, we have $\Mat{W}_l={\rm diag} (\sigma^{-1}_{1l},\ldots,\sigma^{-1}_{Nl})$.
The matrix $\Mat{W}_l$ is useful for pre-whitening the sensor data.
The corresponding whitened sensor data and dictionary matrix are
\alpheqn
\begin{align}
\widetilde{\Vec{y}}_l&=\Mat{W}_l \Vec{y}_l,\\
\widetilde{\Vec{A}}_l&=\Mat{W}_l \Vec{A}~.
 \end{align}
\reseteqn
For known diagonal noise covariance $\Mat\Sigma_{\Vec{n}_l}$ the above means we have to normalize each row in \eqref{eq:lin3} with $\sigma_{nl}$ as then the noise satisfies  $\widetilde{\Vec{n}}_l \sim {\cal CN}(\widetilde{\Vec{n}}_l; \Vec{0}, \Mat{I})$, and thus all entries are identically distributed.

If the noise covariance matrices are not known, they can be estimated as we will show later on. More specifically, at low SNR, the Noise Cases I, II, and III, lead to the noise variance estimates~\eqref{eq:noise-power-estimate-case-I},~\eqref{eq:noiseII}. 
% and~\eqref{eq:noiseIIIL}, respectively.
The corresponding pre-whitening matrices can then be computed as
\begin{align}
\Mat{W}_l \approx 
\left\{
\begin{array}{ll}
\frac{\sqrt{NL}}{\| \Mat{Y}  \|_{\!F}}\,   \Mat{I}& \text{for Case I}\\
\frac{\sqrt{N}}{\| \Mat{y}_l \|_2} \Mat{I}& \text{for Case II}\\
\text{diag}(|y_{1,l}|,\ldots,|y_{N,l}|)^{-1} & \text{for Case III}~,
\end{array}
\right.   \label{eq:w}
\end{align}
leading to
\begin{align}
\widetilde{\Mat{y}}_l \approx % \Mat{\Sigma}_l^{-1/2} =
\left\{
\begin{array}{ll}
\frac{\sqrt{NL}}{\| \Mat{Y}  \|_{\!F}} \Vec{y}_l      	& \text{for Case I}\\
\frac{\sqrt{N}}{\| \Mat{y}_l \|_2} \Vec{y}_l		& \text{for Case II}\\
\left[ y_{1,l}/|y_{1,l}|,\ldots,  y_{N,l}/|y_{N,l}|\right]^{T} & \text{for Case III}~.
\end{array}
\right.   \label{eq:yw}
\end{align}

Empirically, it has been found that  applying pre-whitening to the data $\Vec{y}_l$ only (the dictionary $\Mat{A}$ is non-whitened) effective in finding the strongest DOA.

For Noise Case I, pre-whitening does not play a role and the conventional beamformer is formulated using the power spectrum at DOA $\theta_{m}$
\begin{align}
P_{\text{CBF}}(\theta_{m}) &=\frac{1}{L} \Vec{a}_{m}^{H} \Vec{Y} \Vec{Y}^{H} \Vec{a}_{m}
	= \Vec{a}_{m}^H \Vec{S}_{\Vec{y}} \Vec{a}_{m}~, \label{eq:CBF}
\end{align}
where the sample covariance matrix is
\begin{align}
\Vec{S}_{\Vec{y}} = \frac{1}{L} \Vec{Y} \Vec{Y}^{H}.
 \label{eq:scm}
 \end{align}

For Noise Cases II and III on the other hand, we will work with the pre-whitened data. In those cases, , the conventional beamformer leads to the powerspectra $P_{\text{CBF2}}$ and $P_{\text{CBF-Phase}}$, respectively, which can be formulated as
\begin{align}
P_{\text{CBF*}}(\theta_{m}) &= \frac{1}{L}\Vec{a}_{m}^{H} \Vec{\tilde{Y}} \Vec{\tilde{Y}}^{H} \Vec{a}_{m}
	= \Vec{a}_{m}^H \Vec{S}_{\Vec{\tilde{y}}} \Vec{a}_{m}~,\label{eq:CBF2}
\end{align}
where the whitened sample covariance matrix is
\begin{align}
\Mat{S}_{\Vec{\tilde{y}}} &= \frac{1}{L}\widetilde{\Mat{Y}}\widetilde{\Mat{Y}}^H
	%= \frac{1}{L}\sum_{l=1}^L\Mat{W}_l{\Vec{y}_l}{\Vec{y}_l^H}\Mat{W}_l^H
	= \frac{1}{L}\sum_{l=1}^L{\Vec{y}_l}\Mat{W}_l^2{\Vec{y}_l^H}.
\end{align}
The weighting $\Mat{W}_l$ is given by \eqref{eq:w}. For Case III, only phase is used as can be observed from \eqref{eq:yw}, which results in phase-only processing.
As demonstrated in Section~\ref{sec:OneDoaCBF} this simple pre-whitening can improve the DOA performance at low SNR.

%%%%%%%%%%%%%%%%%%%%%%%%%%%%%%
\section{Source power estimation}\label{sec:src-pwr-estim}

Let us now focus on the SBL algorithm solving~\eqref{eq:TypeIIMaxLikelihood}. The algorithm iterates between the source power estimates $\hat{\Vecgamma}$ derived in this section and the noise variance estimates $\widehat{\Mat{V}}_{\Mat N}$ computed in Sec.~\ref{sec:noise}.
After detailing these estimation procedures, the full algorithm is summarized in Sec.~\ref{sec:sbl-algo}.

We impose the diagonal structure $\mtight \Mat{\Gamma}=\mathop{\mathrm{diag}}
(\Vecgamma)$, in agreement with \eqref{eq:tipping-like}, and form derivatives of \eqref{eq:evidence-L} with respect to the diagonal elements $\gamma_m$,
cf. \cite{Bohme1986}.
Using
\begin{align}
&\frac{\partial \Mat{\Sigma}_{\Vec{y}_l}^{-1}}{\partial \gamma_m} = -\Mat{\Sigma}_{\Vec{y}_l}^{-1} \, \frac{\partial \Mat{\Sigma}_{\Vec{y}_l}}{\partial \gamma_m}  \, \Mat{\Sigma}_{\Vec{y}_l}^{-1}
% &=
% -\Mat{\Sigma}_{\Vec{y}}^{-1} \Mat{A} \, \frac{\partial \Mat{\Gamma}}{\partial \gamma_m}  \, \Mat{A}^H\Mat{\Sigma}_{\Vec{y}}^{-1}\\
%  &= -\Mat{\Sigma}_{\Vec{y}}^{-1} \Mat{A} \, \Vec{e}_m^{}\Vec{e}_m^T  \, \Mat{A}^H\Mat{\Sigma}_{\Vec{y}}^{-1} \\
  = -\Mat{\Sigma}_{\Vec{y}_l}^{-1} \Vec{a}_m^{}\Vec{a}_m^H \Mat{\Sigma}_{\Vec{y}_l}^{-1}, \label{eq:deriv-inverse}
\\
& \frac{\partial\log\det( \Mat{\Sigma}_{\Vec{y}_l})}{\partial\gamma_m}
 = \trace{\Mat{\Sigma}_{\Vec{y}_l}^{-1} \frac{\partial \Mat{\Sigma}_{\Vec{y}_l}}{\partial\gamma_m}} %\\
% &=   \trace{\Mat{\Sigma}_{\Vec{y}}^{-1} \Mat{A} \, \frac{\partial \Mat{\Gamma}}{\partial \gamma_m}  \, \Mat{A}^H} \\
% &=  \trace{\Mat{\Sigma}_{\Vec{y}}^{-1} \Mat{A} \Vec{e}_m^{}\Vec{e}_m^T \Mat{A}^H} \\
% &=  \trace{\Mat{\Sigma}_{\Vec{y}}^{-1} \Vec{a}_m^{}\Vec{a}_m^H} \\
 =   \Vec{a}_m^H   \Mat{\Sigma}_{\Vec{y}_l}^{-1} \Vec{a}_m
 \label{eq:deriv-log-det},
\end{align}
%the derivative of \eqref{eq:evidence-L} is
%\begin{align}
% \frac{\partial\log p(\Mat{Y};\Vecgamma,\sigma^2)}{\partial\gamma_m} %&=
%% \Vec{y}^H\Mat{\Sigma}_{\Vec{y}}^{-1} \Vec{a}_m^{}\Vec{a}_m^H \Mat{\Sigma}_{\Vec{y}}^{-1}\Vec{y} -  \Vec{a}_m^H \Mat{\Sigma}_{\Vec{y}}^{-1} \Vec{a}_m
%%    \nonumber \\
%%    &=  |\Vec{y}^H \Mat{\Sigma}_{\Vec{y}}^{-1}\Vec{a}_m |^2
%%    -  \Vec{a}_m^H \Mat{\Sigma}_{\Vec{y}}^{-1} \Vec{a}_m \\
%%    &=  | \Vec{a}_m^H\Mat{\Sigma}_{\Vec{y}}^{-1} \Vec{y}  |^2
%%    -  \Vec{a}_m^H \Mat{\Sigma}_{\Vec{y}}^{-1} \Vec{a}_m \\
%    &= \frac{1}{\gamma_m^{2}L} \| \Vec{\mu}_m  \|_2^2
%    -  \Vec{a}_m^H \Mat{\Sigma}_{\Vec{y}}^{-1} \Vec{a}_m,
%    \label{eq:score-gamma}
%\end{align}
%where $\mtight \Vec{\mu}_m=\gamma_m\Vec{a}_m^H\Mat{\Sigma}_{\Vec{y}}^{-1}\Mat{Y}$ is the $m$th row of $\Mat{\mu}_{\Mat{X}}$ in \eqref{eq:Wipf-and-Rao-2007}.
the derivative of \eqref{eq:evidence-L} is formulated as
\begin{align}
\!\!\!\! \frac{\partial\log p(\Mat{Y};\!\Vecgamma,\!\Mat{V}_{\Mat N}\!)}{\partial\gamma_m} \!&
 =
\sum_{l=1}^L\! \left( \Vec{a}_m^H\Mat{\Sigma}_{\Vec{y}_l}^{-1} \!\Vec{y}_l^{}\Vec{y}_l^H \Mat{\Sigma}_{\Vec{y}_l}^{-1}\!\Vec{a}_m \!-  \Vec{a}_m^H \Mat{\Sigma}_{\Vec{y}_l}^{-1} \!\Vec{a}_m\right)
    \nonumber \\
&=
\sum_{l=1}^L\! \Vec{a}_m^H \left( \Mat{\Sigma}_{\Vec{y}_l}^{-1} \!\Vec{y}_l^{}\Vec{y}_l^H \Mat{\Sigma}_{\Vec{y}_l}^{-1} \!-   \Mat{\Sigma}_{\Vec{y}_l}^{-1} \right)\Vec{a}_m
    \label{eq:necessary-conditon-v1} \\
&=
\sum_{l=1}^L | \Vec{y}_l^H\Mat{\Sigma}_{\Vec{y}_l}^{-1} \!\Vec{a}_m^{} |^2-  \sum_{l=1}^L\Vec{a}_m^H \Mat{\Sigma}_{\Vec{y}_l}^{-1} \Vec{a}_m.
\label{eq:necessary-conditon-v2}
\end{align}
For the solution to (\ref{eq:TypeIIMaxLikelihood}), we impose the necessary condition $\frac{\partial\log p(\Mat{Y}; {\bf \gamma}, \Mat{V}_{\Mat N})}{\partial\gamma_m}
 =0$.
  To obtain an iterative equation in $\gamma_m$, we multiply the first term in the above equation with the factor $(\frac{\gamma_m^{\rm old}}{\gamma_m^{\rm new}})^{1/b}$ , whereby
\begin{align}
(\frac{\gamma_m^{\rm old}}{\gamma_m^{\rm new}})^{1/b} \! \sum_{l=1}^L | \Vec{y}_l^H\Mat{\Sigma}_{\Vec{y}_l}^{-1} \!\Vec{a}_m^{} |^2-  \sum_{l=1}^L\Vec{a}_m^H \Mat{\Sigma}_{\Vec{y}_l}^{-1} \Vec{a}_m=0~.
 %  \frac{\gamma_m^{2,{\rm old}}}{\gamma_m^2} \nonumber \\
%    &=  |\Vec{y}^H \Mat{\Sigma}_{\Vec{y}}^{-1}\Vec{a}_m |^2
%    -  \Vec{a}_m^H \Mat{\Sigma}_{\Vec{y}}^{-1} \Vec{a}_m \\
%    &=  | \Vec{a}_m^H\Mat{\Sigma}_{\Vec{y}}^{-1} \Vec{y}  |^2
%    -  \Vec{a}_m^H \Mat{\Sigma}_{\Vec{y}}^{-1} \Vec{a}_m \\
%    &= \frac{1}{\gamma_m^{2}L} \| \Vec{\mu}_m  \|_2^2
 %   -  \Vec{a}_m^H \Mat{\Sigma}_{\Vec{y}}^{-1} \Vec{a}_m,
    \label{eq:score-gamma}
\end{align}
Assuming $\gamma_m^{\rm old}$ and $\Mat{\Sigma}_{\Vec{y}_l}$ given (from previous iterations or initialization) and forcing \eqref{eq:score-gamma} to zero, we obtain the following  fixed point iteration \cite{BurdonFaires} for the $\gamma_m$:
\be
%    \gamma_m^{\rm new} = {\frac{1}{\sqrt{L}}\|\Vec{\mu}_m\|_2}/{\sqrt{\Vec{a}_m^H\Mat{\Sigma}_{\Vec{y}}^{-1}\Vec{a}_m}}.
   \gamma_m^{\rm new} = \gamma_m^{\rm old}  \left(\frac{    \sum_{l=1}^L | \Vec{y}_l^H\Mat{\Sigma}_{\Vec{y}_l}^{-1} \Vec{a}_m^{}|^2}{\sum_{l=1}^L{\Vec{a}_m^H\Mat{\Sigma}_{\Vec{y}_l}^{-1}\Vec{a}_m}}\right)^b.
%  \tag{SBL1}
   \label{eq:gamma-update}
\ee
We use $b=0.5$, %It is not clear what value of $b$ to use. Higher $b$-values give faster convergence to an estimate, 
but have not carefully tested for optimal values and this value  depends on many factors such as which noise estimate is used and closeness of the DOAs.
A value of $b = 1$ gives the update equation used
in~\cite{Tipping2001, Wipf2007, nannuru2017} and $b = 0.5$ gives the update equation used in~\cite{gerstoft2016mmv}.

%The work~\cite{Wipf2007} (Eq.\ 18) follows a less direct EM approach to estimate the update~\eqref{eq:gamma-update-EM}:
%\be
%\gamma_m^{\rm new}
%= \frac{(\gamma_m^{\rm old})^2}{{L}} \sum_{l=1}^L | \Vec{y}_l^H \Mat{\Sigma}_{\Vec{y}_l}^{-1} \Vec{a}_m |^2 + (\gamma_m^{\rm old})^2 ,
%%=\frac{(\gamma_m^{\rm old})^2}{{L}}  \| \Vec{Y}^H\Mat{\Sigma}_{\Vec{y}}^{-1} \Vec{a}_m^{}\|_2^2+(\Mat{\Sigma}_{\Vec{x}})_{mm} ,
%%={\frac{1}{L}\|\Vec{\mu}_m\|_2^2}+(\Mat{\Sigma}_{\Vec{x}})_{mm} . \tag{SBL-EM}   \label{eq:gamma-update-EM}
%\ee
%for Noise Cases I and II ($\Mat{\Sigma}_{\Vec{y}_l} = \Mat{\Sigma}_{\Vec{y}}$). 
%
%The sequence of parameter estimates in the EM iteration converges \cite{dempster1977}. However,  convergence is only guaranteed towards a \emph{local} optimum of the marginal log-likelihood \eqref{eq:evidence-L}. We refer to the above update rule as SBL-EM. As seen from simulations %(Sec.\ \ref{sec:example}) both update rules~\eqref{eq:gamma-update} and~\eqref{eq:gamma-update-EM} converge. Guarantees for convergence can be given provided  $\mtight | {\partial \gamma_m^{\rm new}}/{\partial \gamma_{m}}|<1$, but is hard to prove analytically.

%%%%%%%%%%%%%%%%%%
\section{Noise variance estimation}\label{sec:noise}

While there has been more focus on estimating the DOAs or $\Vecgamma$, noise is an important part of the physical system and  a correct estimate is needed for good convergence properties. In SBL, the  noise variance controls the sharpness of the peaks in the $\Vecgamma$ spectrum, with higher noise levels giving broader peaks. Thus, as we optimize the DOAs we expect  the noise levels to decrease and the $\Vecgamma$ spectrum to become sharper.

In this section we estimate the noise variance for the three noise cases in Sec. \ref{sec:noise_models}. In Secs.~\ref{sec:noiseI}--\ref{sec:noiseIIIH}, we will assume the support of $\Vecgamma$ is known. %However, if $\Vecgamma$ is not known, an upper bound for the noise variance can be derived, as done in Sec.~\ref{sec:noiseIIIL}.

\subsection{Noise estimate, Case I}
\label{sec:noiseI}

Under Noise Case I, where $\Mat{{\Sigma}}_{\Vec{n}_l} = {\sigma}^2\Mat{I}_N$ with $\Mat{I}_N$ the identity matrix of size $N$, stochastic maximum likelihood \cite{Liu2012,Boehme1985,Stoica-Nehorai1995} can provide an asymptotically efficient estimate of $\sigma^2$ if the set of active DOAs $\cal M$ is known.

Let  $\mtight \Mat{{\Gamma}}_{\cal M}=\mathop{\mathrm{diag}}(\Vecgamma_{\cal M}^{\rm new})$ be the  covariance matrix of the $K$ active sources obtained above with corresponding active steering matrix $\Mat{A}_{\cal M}$ which maximizes (\ref{eq:evidence-L}). % and $\bar{\sigma}^2$ be the optimal $\sigma^2$.
The corresponding data covariance matrix is
\be
\Mat{{\Sigma}}_{\Vec{y}_l} = {\sigma}^2\Mat{I}_N + \Mat{A}_{\cal M}\Mat{{\Gamma}}_{\cal M}\Mat{A}_{\cal M}^H .
\label{eq:stationary-model-covariance}
\ee
Note that for Noise Case I, the data covariance matrices \eqref{eq:array-covariance} and \eqref{eq:stationary-model-covariance} are identical.
Following \cite{Jaffer1988}, we continue from (\ref{eq:necessary-conditon-v1}),
\begin{align}
\!\!\!\! \frac{\partial\log p(\Mat{Y};\!\Vecgamma,\!\Mat{V}_{\Mat N}\!)}{L \,\,\partial\gamma_m} \!&
 =  \Vec{a}_m^H \left( \Mat{\Sigma}_{\Vec{y}_l}^{-1} \Mat{S}_{\Vec{y}}\Mat{\Sigma}_{\Vec{y}_l}^{-1}  - 
\Mat{\Sigma}_{\Vec{y}_l}^{-1}\right) \Vec{a}_m \\
&=  \Vec{a}_m^H \Mat{\Sigma}_{\Vec{y}_l}^{-1} \left(  \Mat{S}_{\Vec{y}}  - 
\Mat{\Sigma}_{\Vec{y}_l}\right) \Mat{\Sigma}_{\Vec{y}_l}^{-1}\Vec{a}_m = 0,
\label{eq:Jaffer-diagonal-preliminary}
\end{align}
for all active sources ($m\in\mathcal{M}$).
Since $\mathop{\mathrm{range}}(\Mat{\Sigma}_{\Vec{y}_l}^{-1}\Mat{A}_{\cal M}) = \mathop{\mathrm{range}}(\Mat{A}_{\cal M})$, Eq.(\ref{eq:Jaffer-diagonal-preliminary}) simplifies to
\alpheqn
\be
 \Vec{a}_m^H \left(  \Mat{S}_{\Vec{y}}  - 
\Mat{\Sigma}_{\Vec{y}_l}\right) \Vec{a}_m = 0,\quad\forall m\in\mathcal{M}~.
\hspace*{9ex}
\label{eq:Jaffer-diagonal}
\ee
To obtain Jaffer's condition below, we impose the following additional constrains %conditions
\be
 \Vec{a}_m^H \left(  \Mat{S}_{\Vec{y}}  - 
\Mat{\Sigma}_{\Vec{y}_l}\right) \Vec{a}_p = 0,\quad\forall m,p\in\mathcal{M}, m\ne p.
\label{eq:Jaffer-offdiagonal}
\ee
\reseteqn
Together, the conditions (\ref{eq:Jaffer-diagonal}), (\ref{eq:Jaffer-offdiagonal}) give Jaffer's condition (\cite{Jaffer1988}:Eq.(6)), i.e.,
\be
     \Mat{A}_{\cal M}^H \left(  \Mat{S}_{\Vec{y}} - \Mat{{\Sigma}}_{\Vec{y}_l} \right) \Mat{A}_{\cal M} = \Mat{0},
\label{eq:necessary-condition}
\ee
which we enforce at the optimal solution $(\Mat{{\Gamma}}_{\cal M}, {\sigma}^2)$.
%Jaffer's condition follows from allowing unknown correlations among the source signals.
Jaffer's condition follows from allowing arbitrary correlations among the source signals, i.e. when the $\Gamma$ matrix is not restricted to be diagonal. 
%This will maximize the source signal fit to the data and could potentially underestimate the noise $\sigma$. 
Substituting (\ref{eq:stationary-model-covariance}) into (\ref{eq:necessary-condition}) gives
\be
\label{eq:JafferModified}
     \Mat{A}_{\cal M}^H \left(   \Mat{S}_{\Vec{y}} - {\sigma}^2\Mat{I}_N \right) \Mat{A}_{\cal M} =
      \Mat{A}_{\cal M}^H \Mat{A}_{\cal M}\Mat{{\Gamma}}_{\cal M}\Mat{A}_{\cal M}^H \Mat{A}_{\cal M}.
\ee

Let us then define the projection matrix onto the subspace spanned by the active steering vectors
\be
 \P=\A\A^+=\A(\A^H\A)^{-1}\A^H = \P^H = \P^2.
 \label{eq:projection-matrix}
\ee
Left-multiplying~\eqref{eq:JafferModified} with $\mtight \A^{+H}=\A(\A^H\A)^{-1}$ and right-multiplying it with
$\mtight \A^+ = (\A^H\A)^{-1}\A^H$, we obtain
\begin{align}
   \P \Mat{S}_{\Vec{y}} \P^H - \sigma^2 \P \P^H &= \P \A \G \A^H \P^H = \A \G \A^H \nonumber \\
        &= \Mat{\Sigma}_{\Vec{y}_l} - \sigma^2\Mat{I}_N.
%   \P \Mat{S}_{\Vec{y}} \P - \sigma^2 \P &=& \Mat{\Sigma}_{\Vec{y}} - \sigma^2\Mat{I}_N
\end{align}
 Evaluating the trace, using $ \mtight\trace{\P \P^H} = K$ and $\mtight\trace{\P \Mat{S}_{\Vec{y}} \P^H}=\trace{\P \Mat{S}_{\Vec{y}} }$,  gives
%(similar expression is in \cite{Boehme1985,Stoica-Nehorai1995})
\begin{align}
    \sigma^2 &=  \frac{\mathop{\mathrm{tr}}[ \Mat{\Sigma}_{\Vec{y}_l} -  \P \Mat{S}_{\Vec{y}}]}{N-K} =  \frac{\mathop{\mathrm{tr}}[ (\Mat{S}_{\Vec{y}} -  \P \Mat{S}_{\Vec{y}}] +\epsilon}{N-K}  \\
&\approx \frac{ \mathop{\mathrm{tr}}[ (\Mat{I}_N -  \P) \Mat{S}_{\Vec{y}}] }{N-K} = \hat{\sigma}^2~,
    %\label{eq:bar-sigma2}
    \label{eq:noise-power-estimate-case-I}
\end{align}
where we  have defined  $\mtight \epsilon=\mathop{\mathrm{tr}}[\Mat{\Sigma}_{\Vec{y}_l}- \Mat{S}_{\Vec{y}}]$.

The above approximation  motivates the noise power estimate for Noise Case I \eqref{eq:noise-power-estimate-case-I},
%\begin{align}
%    \hat{\sigma}^2 &=  \frac{ \mathop{\mathrm{tr}}[ (\Mat{I}_N -  \P) \Mat{S}_{\Vec{y}}] }{N-K} \label{eq:noise-power-estimate-case-I}
%\end{align}
which is \emph{error-free} if $\mtight \mathop{\mathrm{tr}}[\Mat{\Sigma}_{\Vec{y}}] = \mathop{\mathrm{tr}}[\Mat{S}_{\Vec{y}}]$, % (the whole $ \Mat{S}_{\Vec{y}}$ is not needed to converge).
 \emph{unbiased} because  $ \mathop{\mathrm{E}}[\epsilon] = 0 $, %$\mathop{\mathrm{tr}}[\Mat{\Sigma}_{\Vec{y}}] = \mathop{\mathrm{tr}}[\mathop{\mathrm{E}}(\Mat{S}_{\Vec{y}})]$.
 \emph{consistent} since also its variance tends to zero for $\mtight L\to\infty$ \cite{Kraus1994},
 and \emph{asymptotically efficient} as it approaches the CRLB for $\mtight L\to\infty$ \cite{Bienvenu1983}.
Note that, the Noise Case I estimate  \eqref{eq:noise-power-estimate-case-I} is valid for any number of snapshots, even for just one snapshot.

%%%%%%%%%
\subsection{Noise estimate, Case II}\label{sec:noiseII}

For Noise Case II, where $\Mat\Sigma_l=\sigma^2 _l \Mat I_{N}$, we apply~\eqref{eq:noise-power-estimate-case-I} for each snapshot $l$ individually, leading to
%define the noise power estimate via \eqref{eq:noise-power-estimate-case-I} for each snapshot $l$, individually
\begin{align}
	\hat{\sigma}^2 _l  = \frac{{\rm tr}[ (\Mat{I}_N -  \P)\Vec{y}_l\Vec{y}_l^H] }{N-K} =
\frac{ \| (\Mat{I}_N -  \P)\Vec{y}_l  \|_2^2}{N-K}.
	\label{eq:noiseII}
\end{align}
 Several alternative estimators 
for the noise variance are proposed based on EM~\cite{Wipf2004,Wipf2007,Zhang2011,Tipping2001,Zhang2014}. For a
comparative illustration in Sec.~\ref{sec:noise_section} we will use the iterative noise
estimate from~\cite{Wipf2004}, given by
\begin{align}
(\sigma^{2}_{l})^{\text{new}} \!\!\!\! &= \frac{||\Vec{y}_l - \Mat{A} \widehat{\Vec{x}}_l^{\rm MAP}||^{2}_{F} +
	(\sigma^{2}_{l})^{\text{old}} (M - \sum_{i=1}^{M} \!\!\!\! {\frac{(\Mat\Sigma_{\Vec{x}_l})_{ii}}{\gamma_{i}}}) }{N} ~,
\label{eq:noise_estimate_N1}
\end{align}
where the posterior mean  $ \widehat{\Vec{x}}_l^{\rm MAP}$ and covariance $ \Mat{\Sigma}_{\Vec{x}_l} $  are given in \eqref{eq:x-map-estimate} and \eqref{eq:Sigma_x}.
Empirically, EM noise estimates such as~\eqref{eq:noise_estimate_N1} significantly underestimate the true noise variance in our applications.
% but  \eqref{eq:bar-sigma2} gives a reasonable estimate for $

%%%%%%%%%%
%\subsection{Noise estimate, Case IIb}
% For case IIb, $\Mat\Sigma_l={\rm diag} [\sigma^2 _1,\ldots,\sigma^2_N] $, Thus we estimate $\sigma^2 _n$ by averaging over snapshots
%   \begin{align}
% \sigma^2 _n=    \frac{1}{L}\sum_{l=1}^L \sigma^2 _{nl} .  \label{eq:noiseIIb}
% \end{align}
% where $\sigma^2 _{nl} $ is estimated by \eqref{eq:noiseIIIH}. Due to the averaging, \eqref{eq:noiseIIb} is a good estimate even for low SNR.

%%%%%%%%%%%%
\subsection{Noise estimate, Case III}\label{sec:noiseIIIH}

Let us start from the definition of the noise covariance 
\begin{align}
\Mat{\Sigma}_{{\Vec n}_l} 
&= {\sf E} \left[ ({\Vec{y}}_l - {\bf A} \Vec{x}_l)({\Vec{y}}_l - {\bf A} \Vec{x}_l)^H  \right] \nonumber \\
&= {\sf E} \left[ ({\Vec{y}}_l - {\bf A}_{\mathcal{M}} \Vec{x}_{\mathcal{M},l})({\Vec{y}}_l - {\bf A}_{\mathcal{M}} \Vec{x}_{\mathcal{M},l})^H  \right]~. % \nonumber \\
%&= \Mat{\Sigma}_{{\Vec y}_l} - \Mat{A}_{\mathcal{M}} \Mat{\Gamma}_{\mathcal{M}} \Mat{A}_{\mathcal{M}} ^H \\
%& = (\Mat{I}-\Mat{P})\Mat{\Sigma}_{\Vec{y}_l}  (\Mat{I}-\Mat{P}) 
\label{eq:noiseIIIa}
\end{align}
This motivates plugging-in the single-observation signal  estimate
 $\widehat{\Vec{x}}_{\mathcal{M},l}=\Mat{A}_{\mathcal{M}}^+\Vec{y}_l\in\mathbb{C}^{K}$ for the active (non-zero) entries in $\Vec{x}_l$. This estimate is based on the single observation $\Vec{y}_l$ and the projection matrix \eqref{eq:projection-matrix}, giving the rank-1 estimate
\begin{align}
\widehat{\Mat{\Sigma}}_{{\Vec n}_l} & = (\Mat{I}-\Mat{P})\Vec{y}_l\Vec{y}_l^H (\Mat{I}-\Mat{P}).
\label{eq:noiseIII}
\end{align}
Since the signal estimate $\widehat{\Vec{x}}_{\mathcal{M},l}$ maximizes the estimated signal power, this noise covariance estimate is biased and the noise level is likely underestimated. 

Since we assume the noise independent across sensors, all off-diagonal elements of $\Mat{\Sigma}_{{\Vec n}_l}$ are known to be zero. 
With this constraint in mind, we modify \eqref{eq:noiseIII} as
\begin{align}
\widehat{\Mat{\Sigma}}_{{\Vec n}_l}  &= 
\mathop{\mathrm{diag}} [\sigma^2 _{1l}, \ldots, \sigma^2 _{nl}, \ldots,\sigma^2 _{Nl} ] \nonumber \\
 &= \mathop{\mathrm{diag}}\left[\mathop{\mathrm{diag}}\left[(\Mat{I}-\Mat{P})\Vec{y}_l\Vec{y}_l^H (\Mat{I}-\Mat{P})\right]\right]~.
\label{eq:noiseIIIH}
\end{align}
The estimate~\eqref{eq:noiseIIIH} is demanding as for all the $N \times L$ complex-valued observations in $\Mat{Y}$, we obtain $N \times L$ estimates of the noise variance.
Note that the estimate $\widehat{\Mat{\Sigma}}_{{\Vec n}_l}$  in~\eqref{eq:noiseIII} is not invertible whereas the diagonal constraint   in~\eqref{eq:noiseIIIH} leads to a non-singular estimate of $\Mat{\Sigma}_{{\Vec n}_l}$ with high probability (it is singular only if an element of $\Vec{y}_l$ is $0$).
%$\Mat{\Sigma}_{{\Vec n}_l}$ is singular, but assuming all elements of  $\Vec{y}_l$ is non-zero retaining only the diagonal
% \eqref{eq:noiseIIIH} gives a nonsingular $\Mat{\Sigma}_{{\Vec n}_l}$. 
As a result, the expression for $\Mat{\Sigma}_{{\Vec y}_l}$ that is used for estimating $\Vecgamma$ in~\eqref{eq:gamma-update} is likely invertible.
% {\bf we could use a full $\Sigma^2 _{{\bf n}_l}$  ....! Silly, $N^2\times L$ estimates , but fun. }

On the other hand, an overestimate of the noise is easily obtained by  assuming $\Vec{x}_l=\Vec{0}$ which is equivalent to 
setting $\Mat{P}=\Mat{0}$ in~\eqref{eq:noiseIIIH}, resulting in
\begin{align}
\widehat{\Mat{\Sigma}}_{{\Vec n}_l}  &=
\mathop{\mathrm{diag}}\left[\mathop{\mathrm{diag}}\left[  \Vec{y}_l\Vec{y}_l^H \right]\right]~
\end{align}
or
 \be
 \widehat{\sigma}_{nl}=|y_{nl}|.
 \label{eq:noiseIIIL}
 \ee
This can  be shown be the maximum likelihood estimate for no sources (${\cal M}=0$) or very low power sources.

%%%%%%%%%%%%%%%%%%%%%%%%%
\section{SBL Algorithm}
\label{sec:sbl-algo}

The complete SBL algorithm is summarized in Table~\ref{tab:SBLML2}. The same algorithm is valid for Noise Cases I, II, and III (at high SNR).
Given the observed $\Mat Y$, we iteratively update ${\Mat \Sigma}_{\Vec{y}_{l}}$ \eqref{eq:array-covariance}
by using the current $\Vecgamma$ and $\Mat{\Sigma}_{\Vec{n}_{l}}$. The ${\Mat \Sigma}_{\Vec{y}_{l}}^{-1}$ is
computed directly as the numerical inverse of ${\Mat \Sigma}_{\Vec{y}_{l}}$.
For updating $\mtight \gamma_m$, $m=1,\ldots,M$ we use \eqref{eq:gamma-update}.  %or \eqref{eq:gamma-update-EM} 
For the initialization of $\Vecgamma $ we use the conventional beamformer (CBF) power
\begin{align}
\Vecgamma  	= \rm{diag}[\Vec{A}^H\Vec{S}_{\Vec{{y}}} \Vec{A}].
\end{align}

 Based on the corresponding noise case,~\eqref{eq:noise-power-estimate-case-I}, \eqref{eq:noiseII},
\eqref{eq:noise_estimate_N1}, or \eqref{eq:noiseIIIH} is used to estimate $\Mat{\Sigma}_{\Vec{n}_{l}}$.
%For noise case III at low SNR, the observations are first normalized (Sec. \ref{sec:noiseIIIL}) and the corresponding noise estimate is~\eqref{eq:noiseIII_one_sigma}. 
The noise is initialized using  \eqref{eq:noise-power-estimate-case-I}, \eqref{eq:noiseII}, 
	\eqref{eq:noise_estimate_N1}, or \eqref{eq:noiseIIIH} with $\Vec{P} = \Vec{0}, K = 0$, which provides an over estimate of the noise variance.

The convergence rate $\epsilon$ measures the relative change
in estimated total source power,
\be
\epsilon= \| \Vecgamma^{\rm new}- \Vecgamma^{\rm old} \|_1 \; \big/ \;
\| \Vecgamma^{\rm old}\|_1 ~.
\label{eq:improvement}
\ee
The algorithm stops when $\epsilon\le\epsilon_{\min}$ and the output is the active set
$\mathcal{M}$ (see~\eqref{eq:active-set-gamma}) from which all source parameters are computed.

%%%%%%%%%%%%%%%%
\begin{table}
{\renewcommand{\arraystretch}{1.3}%
\begin{tabular}{rlc} 
\hline \hline

0& Initialize: $\Vecgamma^{\mathrm{new}} = \rm{diag}[\Vec{A}^H\Vec{S}_{\Vec{{y}}} \Vec{A}]$, \\
& \hspace*{2ex} $\Mat{\Sigma}_{\Vec{n}_{l}}^{\mathrm{new}}$ = Eq.\ \eqref{eq:noise-power-estimate-case-I}, \eqref{eq:noiseII}, 
	\eqref{eq:noise_estimate_N1}, or \eqref{eq:noiseIIIH} with $\Vec{P} = \Vec{0}, K = 0$ \\
& \hspace*{2ex} $ \epsilon_{\min}=0.001, \epsilon = 2 \epsilon_{\min}, j = 0, j_{\max}=100$ \\ \hline

1& while $(\epsilon > \epsilon_{\min})$ and $(j<j_{\max})$ & \\

2& \hspace*{3ex}  $\Vecgamma^{\mathrm{old}} \! = \! \Vecgamma^{\mathrm{new}}, \; 
	\Mat{\Gamma} = \mathop{\mathrm{diag}}(\Vecgamma^{\mathrm{old}} ), \;
	\Mat{\Sigma}_{\Vec{n}_{l}}^{\mathrm{old}} = \Mat{\Sigma}_{\Vec{n}_{l}}^{\mathrm{new}}$ & \\

3& \hspace*{3ex}  $\Mat{\Sigma}_{\Vec{y}_{l}} = \Mat{\Sigma}_{\Vec{n}_{l}}^{\mathrm{old}} + 
	\Mat{A} \Mat{\Gamma} \Mat{A}^H$ & \eqref{eq:array-covariance} \\

4& \hspace*{3ex}  $\gamma_m^{\mathrm{new}}$  use \eqref{eq:gamma-update} \\ %or \eqref{eq:gamma-update-EM} & \\

5& \hspace*{3ex}  $ \mathcal{M} \! = \! \{m \in \mathbb{N} |\, \mbox{K largest peaks in} \,\Vecgamma \}\!= \!\{ m_1\ldots m_K \}$  
	\hspace*{-50ex} & \eqref{eq:active-set-gamma} \\

6& \hspace*{3ex}  $\Mat{A}_\mathcal{M} = (a_{m_1},\ldots,a_{m_K}), \, \Vec{P} = \Mat{A}_\mathcal{M} \Mat{A}_\mathcal{M}^{+} $ & \\

7& \hspace*{3ex}  $\Mat{\Sigma}_{\Vec{n}_{l}}^{\mathrm{new}}$ = choose from  \eqref{eq:noise-power-estimate-case-I}, \eqref{eq:noiseII}, 
	\eqref{eq:noise_estimate_N1}, or \eqref{eq:noiseIIIH} & \\

8& \hspace*{3ex}  $\epsilon = \| \Vecgamma^{\mathrm{new}}  -\Vecgamma^{\mathrm{old}} \|_1 / \| \Vecgamma^{\mathrm{old}} \|_1, \quad$ 
	$j \! = \! j+1$ & \eqref{eq:improvement} \\ 

\hline

9& Output: $\mathcal{M}$, $\Vecgamma^{\mathrm{new}}$, $\Mat{\Sigma}_{\Vec{n}_{l}}^{\rm new} $ & \\

\hline\hline
\end{tabular}}
\caption{SBL Algorithm.}
\label{tab:SBLML2}
\end{table}

%%%%%%%%%%%%%%%%%%%%%%

\begin{figure}[tbh] 
\center\includegraphics[width=0.85\columnwidth]{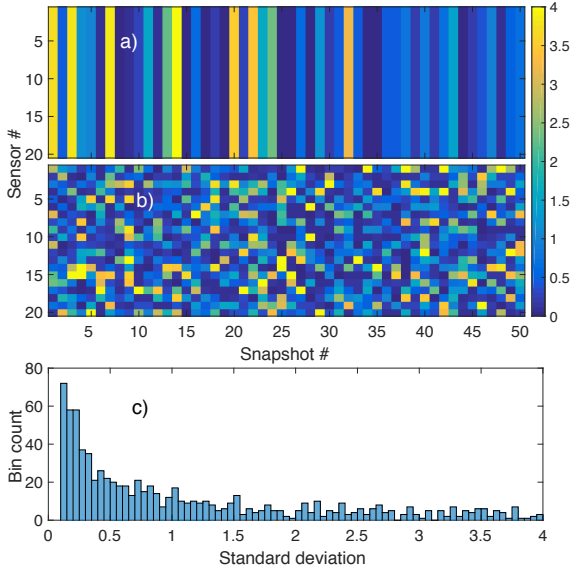}
\caption{Noise  standard deviation matrix $\Mat{V}_{\Mat N}$ in \eqref{eq:noise-std-params} for  Noise Cases  a) II and b)  III.
	c) Histogram of noise for Noise Case III with $\log_{10}  \sigma_{nl}\sim {\cal U}(-1,1)$. 
	In both cases the standard deviation has  mean  1.
	\label{fig:array1noise}}
\end{figure}
\begin{figure}[b] 
\includegraphics[width=0.99\columnwidth]{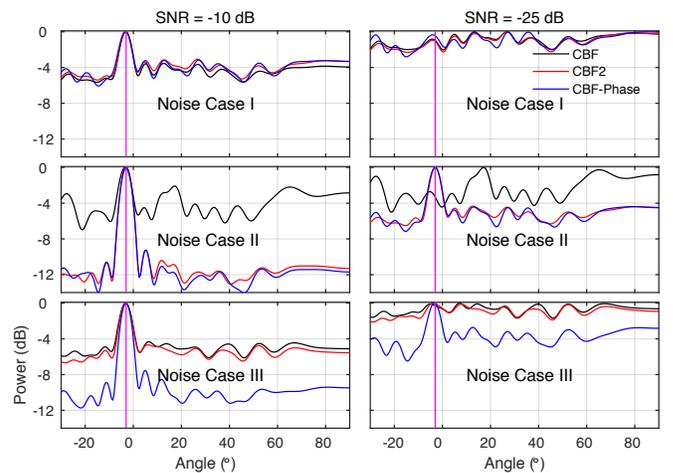}
\caption{ Beam spectra for source at  $-3^\circ$ for SNR $-10$ dB (left) and $-25$ dB (right) for 
	CBF (black), CBF2 (red), CBF-Phase (blue). The noise  standard deviation is a) constant 
	or  heteroscedastic with either b)  $\log_{10} \sigma_l\sim {\cal U}(-1,1)$ or 
	c)  $\log_{10} \sigma_{nl}\sim {\cal U}(-1,1)$. 20 elements and 50 snapshots based 
	on one simulation are used.
         \label{fig:array1}}
\end{figure}

\section{Examples}
\label{se:examples}
When the noise process is stationary and only one source is active, the peak of the CBF spectrum provides the optimal estimate 
of the source location. With heteroscedastic noise this may not hold true. We propose to use a robust CBF 
(Sec.\ref{sec:OneDoaCBF}) or find the DOA with SBL, which takes the heteroscedastic noise into account (Sec.\ \ref{sec:OneDoaSBL}).
With multiple sources and heteroscedastic noise we use SBL (Sec. \ref{sec:ThreeDoaSBL}).

\subsection{Simulation setup}
In the analysis of seismic data  the noise for each snapshot was observed to be log-normal distributed~\cite{Riahi2015}.
Noise  has also been modeled with extreme distributions \cite{Ollila2015}.
In the simulations here, the noise follows a normal-uniform hierarchical model. The noise is complex zero-mean Gaussian with the noise standard deviation uniformly distributed 
over two decades, i.e., $\mtight\log_{10} \sigma_{nl}\sim {\cal U}(-1,1)$, where $\cal U$ is the uniform distribution. 
Three noise cases are simulated: 
(a) Case I: constant noise standard deviation over snapshots and sensors, 
(b) Case II: standard deviation changes across snapshots with $\mtight\log_{10} \sigma_{l} \sim {\cal U}(-1,1)$, and 
(c) Case III: standard deviation changes across both snapshots and sensors with $\mtight\log_{10} \sigma_{nl} \sim {\cal U}(-1,1)$.

A realization 
of the noise standard deviation is shown for Noise Cases II (Fig.~\ref{fig:array1noise}a) and  III (Fig.~\ref{fig:array1noise}b) 
for $\mtight N=20$ sensors and $\mtight L=50$ snapshots. The corresponding histogram   is  presented in 
Fig.~\ref{fig:array1noise}c.

\subsection{Single DOA with CBF and MUSIC}
\label{sec:OneDoaCBF}
 
The single source is located at $-3^{\circ}$. The complex source amplitude is stochastic and 
there is additive heteroscedastic  Gaussian noise with SNR variation from $-40$ to $0$ dB.
The sensor array has $\mtight N = 20$ elements with half wavelength spacing. We process
$\mtight L = 50$ snapshots. The angle space $[-90,90]^{\circ}$ is divided into a  $0.5^{\circ}$ grid  ($\mtight M = 360$). The single-snapshot array signal-to-noise ratio (SNR) is
\be
\mtight \mathrm{SNR} =10 \log_{10}[%\frac
{ \mathop{\mathsf{E}}\left\lbrace\lVert\Mat{A}\Vec{x}_l\rVert_{2}^{2}\right\rbrace}/{ \mathop{\mathsf{E}}\left\lbrace\lVert\Vec{n}_l\rVert_{2}^{2}\right\rbrace} ].
\label{eq:ArraySNR}
\ee

We first compute the beam spectra using CBF  \eqref{eq:CBF}, CBF2 and CBF-Phase  both \eqref{eq:CBF2} but with different weighting  \eqref{eq:yw}. 
 When the noise is homoscedastic 
(constant standard deviation), the beam spectra for the three processors behave similarly (first row in Fig.\ \ref{fig:array1}). For heteroscedastic noise CBF2 and CBF-Phase give  much better discrimination between 
signal to noise, see Fig.\ \ref{fig:array1} middle (Noise Case II) and bottom row (Noise Case III).

The root mean squared error (RMSE)  of the DOA estimates   over $ \mtight N_{\text{sim}} = 100$ runs with random noise realizations is used for evaluating the algorithm
\begin{align}
%\text{RMSE} &= \sqrt{\sum_{n=1}^{N_{sim}}\frac{(\hat{\theta}_n - \theta_{0})^2}{N_{sim}}} \,,
\text{RMSE} &= \sqrt{\sum_{n=1}^{N_{\rm sim}} \sum_{i=1}^{K} \frac{(\hat{\theta}^n_i - \theta^0_i)^2}{N_{\rm sim} K}} \,,
\end{align}
where $\theta^0_i$ is the true DOA and $\hat{\theta}^n_{i}$ is the estimated DOA for the $i$th source
when $K$ sources are present. 
%For each source, deviations $\hat{\theta}^n_i - \theta^0_i$ larger than 30$\circ$ are truncated.

\begin{figure}[t]
\center\includegraphics[width=0.85\columnwidth]{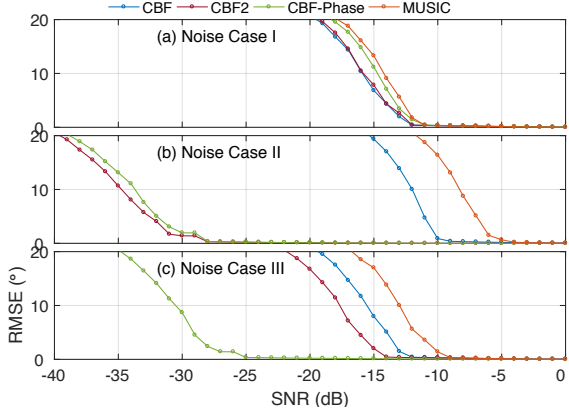}  
\caption{ Single source at $\theta=-3^\circ$: Array RMSE using beamforming with pre-whitening for
         Noise Cases (a)  I, (b)  II, and (c)  III.   Each noise case is solved with CBF, CBF2, CBF-Phase, and MUSIC.
          \label{fig:array1SNR}}
\end{figure}

The SNR curves (Fig.\ \ref{fig:array1SNR}) demonstrate increased robustness of CBF2 and CBF-Phase, failing 20 dB later. Due to the heteroscedastic 
noise, MUSIC performs worse than CBF for a single source. CBF-Phase diverges at an SNR 15--20 dB later 
than CBF for Noise Cases II and III.

\begin{figure}[tb] 
 	 \center\includegraphics[width=0.85\columnwidth]{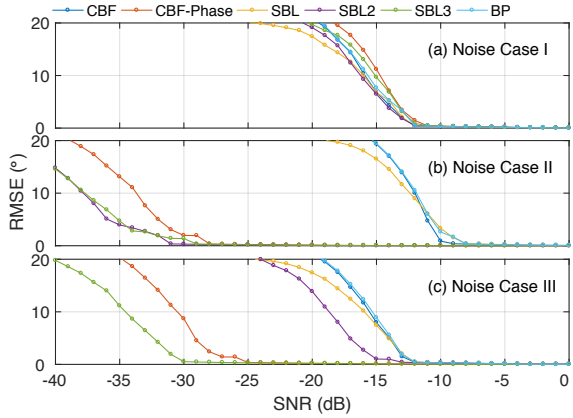}  
        \caption{Single source at $-3^{\circ}$: RMSE vs.\ SNR for DOA estimation using various 
        algorithms for Noise Cases (a)  I, (b)  II, and (c)  III.}
        \label{fig:SingleSource_03deg}
\end{figure}

%%%%%%%%%%%%%%%
\subsection{Single DOA estimation with SBL}
\label{sec:OneDoaSBL}

We use the following SBL methods, with $\Vecgamma$ update   \eqref{eq:gamma-update} and
%\begin{itemize}
\begin{description}
\item[\hspace{-2ex}SBL:] \hspace{-2ex}Solves Case I using standard SBL,
 with $\sigma$ from \eqref{eq:noise-power-estimate-case-I}.
\item[\hspace{-2ex}SBL2:] \hspace{-2ex}Solves  Case II,
with $\sigma_l$ for each snapshot from  \eqref{eq:noiseII}.
\item[\hspace{-2ex}SBL3:]\hspace{-2ex} Solves  Case III,
with  $\sigma_{nl}$ from  \eqref{eq:noiseIIIH}.
\end{description}%\end{itemize}
 In addition to 
these methods we also use basis pursuit (BP)  as implemented in~\cite{Gerstoft2015}. 

For Noise Case I, all the methods fail near the same SNR of $-12.5$ dB, 
Fig.~\ref{fig:SingleSource_03deg}a. When the noise is heteroscedastic across snapshots, 
CBF, BP, and SBL fail early. Since both SBL2 and SBL3 correctly model the noise, they perform
the best for Noise Case II, Fig.~\ref{fig:SingleSource_03deg}b. CBF-Phase 
 also performs well. For heteroscedastic Noise Case III, SBL3 fails the last and as before CBF-Phase  also performs well. SBL3 
fails 15 dB later than any other method. This demonstrates the usefulness 
of accurate noise models in DOA processing.

\begin{figure}[tb] 
 	\center\includegraphics[width=0.85\columnwidth]{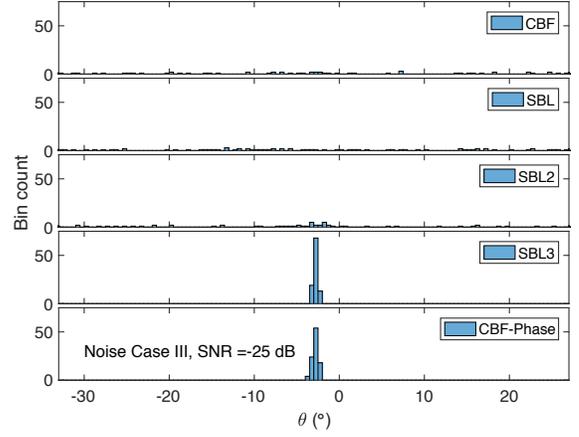}  
        \caption{Single source at $-3^{\circ}$: Histogram of the peak location for Noise Case III at SNR  $-25$ dB.}
        \label{fig:HistCase3}
\end{figure}

The histograms (Fig.~\ref{fig:HistCase3}) of the DOA (location of peak in power 
spectrum)  at  SNR $-25$ dB for Noise Case III demonstrate that when the heteroscedastic  noise is accounted for,  the histograms 
 concentrate  near the true DOA  ($\theta = -3^{\circ}$).
 
\begin{figure}[tb]
\center\includegraphics[width = 0.9\columnwidth]{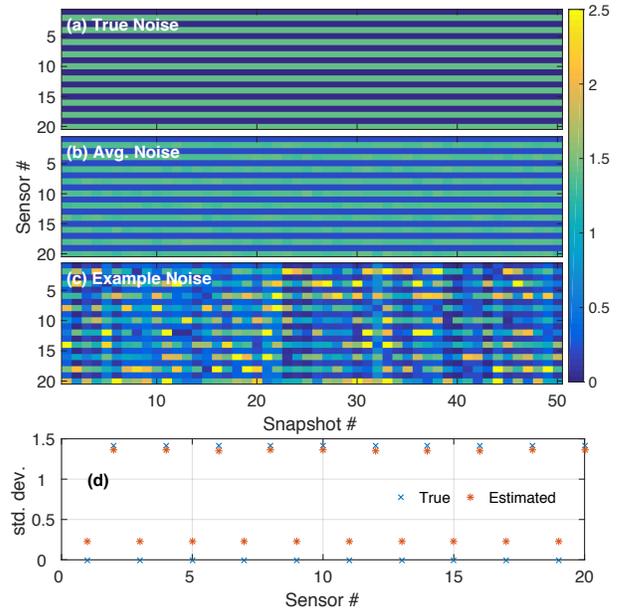}
\caption{Single source at $-3^{\circ}$, SNR=$0$ dB: SBL3 noise standard deviation matrix $\Mat{V}_{\Mat N}$ in \eqref{eq:noise-std-params}: 
(a) true standard deviations, 
(b) average estimated  (500 simulations) standard deviations, 
(c) a typical estimate of standard deviations, and 
(d) average standard deviations across simulations and snapshots.
\label{fig:SingleSourceNoisePattern}}
\end{figure}

An example statistic of the heteroscedastic noise standard deviation is shown in 
Fig.~\ref{fig:SingleSourceNoisePattern}. The standard deviation for each 
 sensor is either $0$ or $\sqrt{2}$. SBL3  estimates
the standard deviation from~\eqref{eq:noiseIIIH}. Average noise in Fig.~\ref{fig:SingleSourceNoisePattern}b resembles well the true noise (Fig.~\ref{fig:SingleSourceNoisePattern}a) whereas the sample 
estimate  (Fig.~\ref{fig:SingleSourceNoisePattern}c) has high variance. 
Fig.~\ref{fig:SingleSourceNoisePattern}d plots the mean across simulations and 
snapshots of the estimated  standard deviation. On average, the noise 
estimate is close to the true noise.

\begin{figure}[tb] 
  \center	\includegraphics[width=0.85\columnwidth]{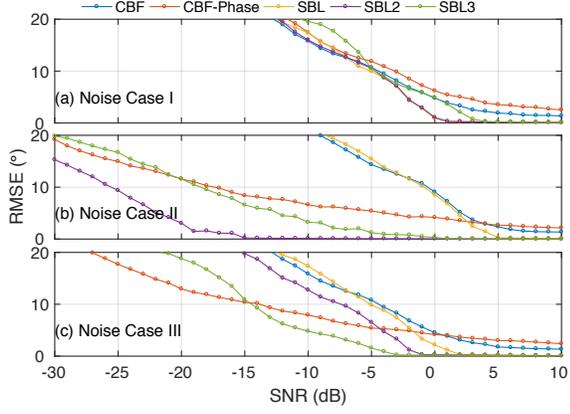}  
        \caption{RMSE vs.\ SNR with the three sources  at $\{-3^{\circ}, 2^{\circ}, 50^{\circ}\}$
        and power $\{10,22,20\}$ dB.}
        \label{fig:ThreeSources}
\end{figure}
\begin{figure}[tb] 
 	\center\includegraphics[width=0.85\columnwidth]{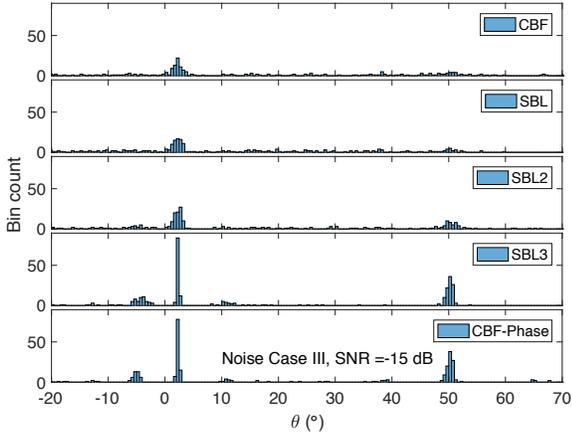}  
        \caption{Three sources at $\{-3^{\circ}, 2^{\circ}, 50^{\circ}\}$: Histogram of the top three peak locations 
	for Noise Case III at SNR  $-15$ dB.}
        \label{fig:HistCase3_3s}
\end{figure}
\begin{figure}[tb] 
  \center	\includegraphics[width=0.85\columnwidth]{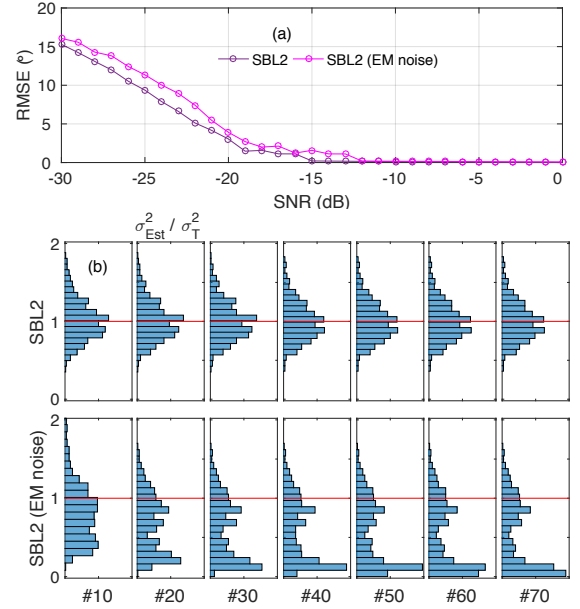}  
        \caption{Three sources at $\{-3^{\circ}, 2^{\circ}, 50^{\circ}\}$ (Noise Case II): 
        		(a) RMSE vs.\ SNR  for SBL2 with two noise estimates.
		(b) Evolution of histogram of noise variance $\sigma_{\text{Est}}^2/\sigma_T^2$ for SBL2 \eqref{eq:noiseII} and SBL2 with EM noise estimate \eqref{eq:noise_estimate_N1} with iterations
		at SNR  $-10$ dB.}
        \label{fig:3S_Noise}
\end{figure}

%%%%%%%
\subsection{Three DOA estimation with SBL}
\label{sec:ThreeDoaSBL}  

Now consider three  sources located at $[-3, 2, 50]^{\circ}$ with power
$[10, 22, 20]$ dB, see  Fig.\ \ref{fig:ThreeSources}.   Relative to the single source case, the CBF-Phase 
performs significantly worse than SBL3, as the  sources at $-3^{\circ}$ and $2^{\circ}$ are both in the CBF main lobe.

The localization ability of the algorithms can also be gauged from the histogram 
of the top three peaks, see Fig.~\ref{fig:HistCase3_3s}. Since SBL3 accounts for the heteroscedastic Noise Case III, 
its histogram is  concentrated near the true DOAs.

\subsubsection{Noise estimate convergence}
\label{sec:noise_section}

The performance of the SBL methods is strongly related to the accuracy  of the noise  estimates. 
For the $l$th snapshot, the true noise power ${\sigma}^2_{l,T}$  (Noise Case II) is
%{\sigma}^2_T=\mathop{\mathsf{E}}[ \| \Vec{N} \|_{\mathcal{F}}^2]/L/N = 10^{-{\rm SNR}/10} \; \mathop{\mathsf{E}} \|\Mat{AX}\|^2_{\mathcal{F}}/L/N.
\be
{\sigma}^2_{l,T}=\mathop{\mathsf{E}}[ \| \Vec{n}_{l} \|_{2}^2]/N = 10^{-{\rm SNR}/10} \; \mathop{\mathsf{E}} [\|\Mat{A}\Vec{x}_{l}\|^2_{2}]/N.
\label{eq:sigma_exp}
\ee 
The estimated $({\sigma}^2_{l})^{\rm new}$ \eqref{eq:noiseII} deviates from ${\sigma}^2_{l,T}$ \eqref{eq:sigma_exp} randomly.
We also consider the noise estimate~\eqref{eq:noise_estimate_N1} based on the EM method.
Fig.~\ref{fig:3S_Noise} compares  the two  noise estimates for noise   generated with $\log \sigma_{l} \sim {\cal U}(-1,1)$. 
The evolution of the histograms of the relative noise variance $\sigma_{\text{Est}}^2/\sigma_T^2$ with iterations is  in 
Fig.~\ref{fig:3S_Noise}b. 
The mean of the ratio of $\sigma_{\text{Est}}^2/\sigma_T^2$ is close to 1 for SBL2 but is much lower for the EM noise estimate \eqref{eq:noise_estimate_N1}, this likely causes the SBL2 (EM noise, \eqref{eq:noise_estimate_N1}) to fails 5 dB earlier.

\begin{figure}[tb]
 \center
  	\center\includegraphics[width=0.85\columnwidth]{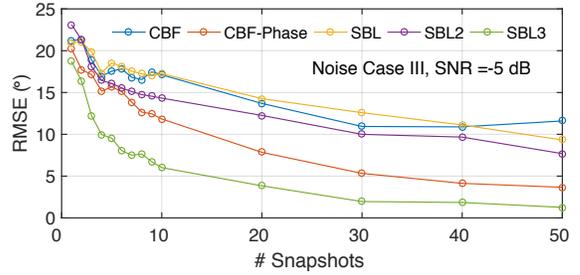}  
        \caption{Three sources at $\{-3^{\circ}, 2^{\circ}, 50^{\circ}\}$ (Noise Case III): RMSE vs.\ Number of snapshots
        with  SNR  $-5$ dB.}
        \label{fig:SingleSourceSnapshots}
\end{figure}
\begin{figure}[tb]
 \center	\includegraphics[width=0.85\columnwidth]{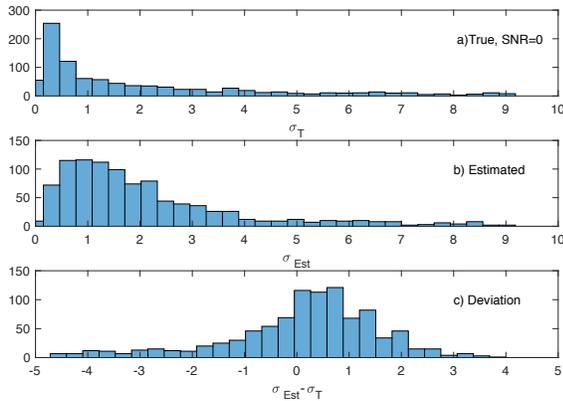}  
  \caption{ Three sources at $\{-3^{\circ}, 2^{\circ}, 50^{\circ}\}$ (Noise Case III): Histogram of noise for a realization for SBL3 with 50 snapshots and 20 sensors (1000 observations) with SNR  $0$ dB. a) True  $\sigma_T$, b) estimated $\sigma_{\text{Est}}$, and c) deviation $\mtight\sigma_{\text{Est}}-\sigma_T$.}
        \label{fig:HistogramSigmaIII}
\end{figure}

\subsubsection{Number of snapshots}
The RMSE versus  number of snapshots
for Noise Case III (Fig.~\ref{fig:SingleSourceSnapshots})  shows that SBL3 is most accurate  with  CBF-Phase following.

\subsubsection{Noise distribution}
For Noise Case III, we are just using one observation to estimate the standard deviation for SBL3 \eqref{eq:noiseIIIH}. Thus the estimates are not  good, see Fig.~\ref{fig:HistogramSigmaIII}.
The true noise standard deviation is generated from $\mtight\log_{10} \sigma_{T,nl}\sim {\cal U}(-1,1)$, see Fig. \ref{fig:HistogramSigmaIII}a. The  distribution of the deviation from the true standard deviation $\mtight\sigma_{\text{Est}}-\sigma_T$ (Fig. \ref{fig:HistogramSigmaIII}c) is well-centered (mean 0.007).
% while the distribution is centered between  0--1 it has a large negative tail. 

\section{Conclusion}

Likelihood based methods for single and multiple DOA estimation from long-term observations corrupted by nonstationary additive noise are discussed.
In such a setting, the DOA estimators  for a stationary noise model perform poorly.
Therefore a heteroscedastic Gaussian noise model is introduced where the noise variance varies across sensors and snapshots.
We develop Sparse Bayesian Learning (SBL) approaches to estimate source powers, source DOAs, and the noise variance parameters.

Simulations show that normalizing the array data magnitude (such that only the phase information is retained) is simple and useful for estimating a single source in  heteroscedastic Gaussian noise. % maybe this can also be shown analytically: does the likelihood depend only on the data phase?
For the estimation of (multiple) closely spaced sources, a problem specific SBL approach gives a much lower RMSE in the DOA estimates.

\bibliographystyle{unsrt}
\bibliography{PhaseREFJuly}

\begin{thebibliography}{10}

\bibitem{Murphy2012}
K.~P. Murphy.
\newblock {\em Machine learning: a probabilistic perspective}.
\newblock MIT press, 2012.

\bibitem{Bishop2006}
C.~M. Bishop.
\newblock {\em Pattern recognition and machine learning}.
\newblock Springer, 2006.

\bibitem{huber2011}
P.~J. Huber.
\newblock {\em Robust statistics}.
\newblock Springer, 2011.

\bibitem{maronna2006}
R.~Maronna, D.~Martin, and V.~Yohai.
\newblock {\em Robust statistics}.
\newblock John Wiley, Chichester., 2006.

\bibitem{engle1982}
R.~F. Engle.
\newblock Autoregressive conditional heteroscedasticity with estimates of the
  variance of united kingdom inflation.
\newblock {\em Econometrica}, pages 987--1007, 1982.

\bibitem{thai2014}
T.H. Thai, R.~Cogranne, and F~Retraint.
\newblock Camera model identification based on the heteroscedastic noise model.
\newblock {\em IEEE Trans. Image Proc.}, 23(1):250--263, 2014.

\bibitem{viberg1997}
M.~Viberg, P.~Stoica, and B.~Ottersten.
\newblock Maximum likelihood array processing in spatially correlated noise
  fields using parameterized signals.
\newblock {\em IEEE Trans Signal Proc.}, 45(4):996--1004, 1997.

\bibitem{chen2008}
C.~E. Chen, F.~Lorenzelli, R.E Hudson, and K.~Yao.
\newblock Stochastic maximum-likelihood doa estimation in the presence of
  unknown nonuniform noise.
\newblock {\em IEEE Trans. Signal Proc.}, 56(7):3038--3044, 2008.

\bibitem{li2011}
T~Li and A.~Nehorai.
\newblock Maximum likelihood direction finding in spatially colored noise
  fields using sparse sensor arrays.
\newblock {\em IEEE Trans Signal Proc.}, 59(3):1048--1062, 2011.

\bibitem{cox1973line}
H.~Cox.
\newblock Line array performance when the signal coherence is spatially
  dependent.
\newblock {\em J Acoust. Soc. Am.}, 54(6):1743--1746, 1973.

\bibitem{paulraj1988}
A.~Paulraj and T.~Kailath.
\newblock Direction of arrival estimation by eigenstructure methods with
  imperfect spatial coherence of wave fronts.
\newblock {\em J Acoust. Soc. Am.}, 83(3):1034--1040, 1988.

\bibitem{lefort2017}
R.~Lefort, R.~Emmeti{\`e}re, S.~Bourmani, G.~Real, and A.~Dr{\'e}meau.
\newblock Sub-antenna processing for coherence loss in underwater direction of
  arrival estimation.
\newblock {\em J Acoust. Soc. Am.}, 142(4):2143--2154, 2017.

\bibitem{law2006}
N.M Law, C.~D. Mackay, and J.E Baldwin.
\newblock Lucky imaging: high angular resolution imaging in the visible from
  the ground.
\newblock {\em Astronomy \& Astrophysics}, 446(2):739--745, 2006.

\bibitem{ge2016}
H.~Ge and I.~P. Kirsteins.
\newblock Lucky ranging with towed arrays in underwater environments subject to
  non-stationary spatial coherence loss.
\newblock In {\em IEEE Intl. Conf. Acoust., Speech and Signal Proc. (ICASSP)},
  pages 3156--3160, 2016.

\bibitem{gerstoft2006}
P.~Gerstoft, M.~C. Fehler, and K.~G. Sabra.
\newblock When katrina hit california.
\newblock {\em Geophys. Res. Lett.}, 33(17), 2006.

\bibitem{gerstoft2016}
P.~Gerstoft and P.~D. Bromirski.
\newblock ``{W}eather bomb'' induced seismic signals.
\newblock {\em Science}, 353(6302):869--870, 2016.

\bibitem{roux2005}
P.~Roux, K.~G. Sabra, P.~Gerstoft, W.~A. Kuperman, and M.~C. Fehler.
\newblock P-waves from cross-correlation of seismic noise.
\newblock {\em Geophys. Res. Lett.}, 32(19), 2005.

\bibitem{harmon2008}
N.~Harmon, P.~Gerstoft, C.~A. Rychert, G.~A. Abers, M.~Salas~de La~Cruz, and
  K.~M. Fischer.
\newblock Phase velocities from seismic noise using beamforming and cross
  correlation in {C}osta {R}ica and {N}icaragua.
\newblock {\em Geophy. Res. Lett.}, 35(19), 2008.

\bibitem{landes2010}
M.~Land{\`e}s, F.~Hubans, N.~M Shapiro, A.~Paul, and M.~Campillo.
\newblock Origin of deep ocean microseisms by using teleseismic body waves.
\newblock {\em J. Geophys Res: Solid Earth}, 115(B5), 2010.

\bibitem{zhan2010}
Z.~Zhan, S.~Ni, D.~V Helmberger, and R.~W. Clayton.
\newblock Retrieval of moho-reflected shear wave arrivals from ambient seismic
  noise.
\newblock {\em Geophys. J. Int.}, 182(1):408--420, 2010.

\bibitem{weemstra2014}
C.~Weemstra, W.~Westra, R.~Snieder, and L.~Boschi.
\newblock On estimating attenuation from the amplitude of the spectrally
  whitened ambient seismic field.
\newblock {\em Geophys. J. Int.}, 197(3):1770--1788, Apr 2014.

\bibitem{gerstoft2008}
P.~Gerstoft, W.~S. Hodgkiss, M.~Siderius, C.-F. Huang, and C.H. Harrison.
\newblock Passive fathometer processing.
\newblock {\em J. Acoust. Soc. Am.}, 123(3):1297--1305, 2008.

\bibitem{sabra2004}
K.G. Sabra and D.R. Dowling.
\newblock Blind deconvolution in ocean waveguides using artificial time
  reversal.
\newblock {\em J. Acoust. Soc. Am.}, 116(1):262--271, 2004.

\bibitem{sabra2010}
K.G. Sabra, H-C. Song, and D.R. Dowling.
\newblock Ray-based blind deconvolution in ocean sound channels.
\newblock {\em J.\ Acoust.\ Soc.\ Am.}, 127(2):EL42--EL47, 2010.

\bibitem{schwartz2014}
O.~Schwartz and S.~Gannot.
\newblock Speaker tracking using recursive em algorithms.
\newblock {\em IEEE Trans. Audio, Speech, Language Proc.}, 22(2):392--402,
  2014.

\bibitem{dorfan2015}
Y~Dorfan and S~Gannot.
\newblock Tree-based recursive expectation-maximization algorithm for
  localization of acoustic sources.
\newblock {\em IEEE Trans. Audio, Speech and Lang. Proc.}, 23(10):1692--1703,
  2015.

\bibitem{VanTreesBook}
H.L. Van~Trees.
\newblock {\em Optimum Array Processing}, chapter 1--10.
\newblock Wiley-Interscience, New York, 2002.

\bibitem{MalioutovDOA:2005}
D.~Malioutov, M.~{\c{C}}etin, and A.~S. Willsky.
\newblock A sparse signal reconstruction perspective for source localization
  with sensor arrays.
\newblock {\em IEEE Trans. Signal Process.}, \textbf{53}(8):3010--3022, 2005.

\bibitem{XenakiCS:2014}
A.~Xenaki, P.~Gerstoft, and K.~Mosegaard.
\newblock Compressive beamforming.
\newblock {\em J. Acoust. Soc. Am.}, \textbf{136}(1):260--271, 2014.

\bibitem{Wipf2007}
D.~P. Wipf and B.D. Rao.
\newblock An empirical {B}ayesian strategy for solving the simultaneous sparse
  approximation problem.
\newblock {\em IEEE Trans. Signal Proc.}, \textbf{55}(7):3704--3716, 2007.

\bibitem{Gerstoft2015}
P.~Gerstoft, A.~Xenaki, and C.F. Mecklenbr{\"a}uker.
\newblock Multiple and single snapshot compressive beamforming.
\newblock {\em J. Acoust. Soc. Am.}, 138(4):2003--2014, 2015.

\bibitem{gerstoft2016mmv}
P.~Gerstoft, C.~F. Mecklenbr{\"a}uker, A.~Xenaki, and S.~Nannuru.
\newblock Multisnapshot sparse {B}ayesian learning for {DOA}.
\newblock {\em IEEE Signal Proc. Lett.}, 23(10):1469--1473, 2016.

\bibitem{nannuru2017}
S.~Nannuru, K.L Gemba, P.~Gerstoft, W.S. Hodgkiss, and C.F. Mecklenbr{\"a}uker.
\newblock Sparse {B}ayesian learning with uncertainty models and multiple
  dictionaries.
\newblock {\em arXiv:1704.00436}, 2017.

\bibitem{Stoica2012}
P.~Stoica and P.~Babu.
\newblock {SPICE} and {LIKES}: Two hyperparameter-free methods for
  sparse-parameter estimation.
\newblock {\em Signal Proc.}, 92(7):1580--1590, 2012.

\bibitem{Wipf2007beam}
D.~P. Wipf and S.~Nagarajan.
\newblock Beamforming using the relevance vector machine.
\newblock In {\em Proc. 24th Int. Conf.\ Machine Learning}, New York, NY, USA,
  2007.

\bibitem{Wipf2004}
D.~P. Wipf and B.D. Rao.
\newblock Sparse {B}ayesian learning for basis selection.
\newblock {\em IEEE Trans. Signal Proc}, 52(8):2153--2164, 2004.

\bibitem{Zhang2011}
Z.~Zhang and B.~D Rao.
\newblock Sparse signal recovery with temporally correlated source vectors
  using sparse {B}ayesian learning.
\newblock {\em IEEE J Sel. Topics Signal Proc.,}, 5(5):912--926, 2011.

\bibitem{Liu2012}
Z.-M. Liu, Z.-T. Huang, and Y.-Y. Zhou.
\newblock An efficient maximum likelihood method for direction-of-arrival
  estimation via sparse {B}ayesian learning.
\newblock {\em IEEE Trans.\ Wireless Comm.}, 11(10):1--11, Oct.\ 2012.

\bibitem{Zhang2016}
JA. Zhang, Z.~Chen, P.~Cheng, and X.~Huang.
\newblock Multiple-measurement vector based implementation for
  single-measurement vector sparse {B}ayesian learning with reduced complexity.
\newblock {\em Signal Proc.}, 118:153--158, 2016.

\bibitem{Giri2016}
R.~Giri and B.~Rao.
\newblock Type {I} and type {II} {B}ayesian methods for sparse signal recovery
  using scale mixtures.
\newblock {\em IEEE Trans Signal Proc. Signal Proc.}, 64(13):3418--3428, July
  2016.

\bibitem{Boehme1985}
J.F. B{\"o}hme.
\newblock Source-parameter estimation by approximate maximum likelihood and
  nonlinear regression.
\newblock {\em IEEE J. Oceanic Eng.}, 10(3):206--212, 1985.

\bibitem{Jaffer1988}
A.G. Jaffer.
\newblock Maximum likelihood direction finding of stochastic sources: A
  separable solution.
\newblock In {\em IEEE Int. Conf. on Acoust., Speech, and Sig. Proc.
  {(ICASSP-88)}}, volume~5, pages 2893--2896, 1988.

\bibitem{Stoica-Nehorai1995}
P.~Stoica and A.~Nehorai.
\newblock On the concentrated stochastic likelihood function in array
  processing.
\newblock {\em Circuits Syst.\ Signal Proc.}, 14(5):669--674, 1995.

\bibitem{Bohme1986}
Johann~F B{\"o}hme.
\newblock Estimation of spectral parameters of correlated signals in
  wavefields.
\newblock {\em Signal Proc.}, 11(4):329--337, 1986.

\bibitem{Krim1996}
H.~Krim and M.~Viberg.
\newblock Two decades of array signal processing research: the parametric
  approach.
\newblock {\em IEEE Signal Proc. Mag.}, 13(4):67--94, 1996.

\bibitem{Stoica1996}
P.~Stoica, B.~Ottersten, M.~Viberg, and R.L. Moses.
\newblock Maximum likelihood array processing for stochastic coherent sources.
\newblock {\em IEEE Trans. Signal Proc.}, 44(1):96--105, 1996.

\bibitem{Tipping2001}
M.~E. Tipping.
\newblock Sparse {B}ayesian learning and the relevance vector machine.
\newblock {\em J. Machine Learning Research}, 1:211--244, 2001.

\bibitem{BurdonFaires}
R.L. Burden, J.D. Faires, and A.M. Burden.
\newblock {\em Numerical analysis}.
\newblock Cengage Learning, 2016.
\newblock (chapter 2.2).

\bibitem{Kraus1994}
D.~Kraus.
\newblock {\em {A}pproximative {M}aximum-{L}ikelihood-{S}ch{\"a}tzung {u}nd
  {v}erwandte {V}erfahren {z}ur {O}rtung {u}nd {S}ignalsch{\"a}tzung {m}it
  {S}ensorgruppen}.
\newblock Shaker Verlag, Aachen, Germany, 1993.

\bibitem{Bienvenu1983}
G.~Bienvenu and L.~Kopp.
\newblock Optimality of high resolution array processing using the eigensystem
  approach.
\newblock {\em IEEE Trans on Acous., Speech, Signal Proc.}, 31(5):1235--1248,
  {Oct} 1983.

\bibitem{Zhang2014}
Z~Zhang, T-P Jung, S.~Makeig, Zhouyue P, and BD. Rao.
\newblock Spatiotemporal sparse {B}ayesian learning with applications to
  compressed sensing of multichannel physiological signals.
\newblock {\em IEEE Trans. Neural Syst. and Rehab. Eng.}, 22(6):1186--1197, Nov
  2014.

\bibitem{Riahi2015}
N.~Riahi and P.~Gerstoft.
\newblock The seismic traffic footprint: Tracking trains, aircraft, and cars
  seismically.
\newblock {\em Geophys. Res. Lett.}, 42(8):2674--2681, 2015.

\bibitem{Ollila2015}
E.~Ollila.
\newblock Multichannel sparse recovery of complex-valued signals using
  {H}uber's criterion.
\newblock In {\em $3^{\mathrm{rd}}$ Int. Workshop on Compressed Sensing Theory
  and Appl. to Radar, Sonar, and Remote Sensing}, {Pisa, Italy}, {June} 2015.

\end{thebibliography}
\end{document}